\documentclass[12pt,preprint]{aastex}
\usepackage{graphicx,times}             
\usepackage{subfigure}
\usepackage{ulem}
\shorttitle{Discovery of $\gamma$-ray emission of S-NLS1s}

\shortauthors{Liao et al.}

\begin{document}

\title{Discovery of $\gamma$-ray emission from steep radio spectrum NLS1s}

\author{Neng-Hui Liao\altaffilmark{1}, Yun-Feng Liang\altaffilmark{1,2}, Shan-Shan Weng\altaffilmark{3}, Marco Berton\altaffilmark{4,5}, Min-Feng Gu\altaffilmark{6}, Yi-Zhong Fan\altaffilmark{1}}
\affil{$^1$ Key Laboratory of Dark Matter and Space Astronomy, Purple Mountain Observatory, Chinese Academy of Sciences, Nanjing 210008, China}
\affil{$^2$ University of Chinese Academy of Sciences, Yuquan Road 19, Beijing 100049, China}
\affil{$^3$ Department of Physics, Nanjing Normal University, Nanjing, China}
\affil{$^4$ Dipartimento di Fisica e Astronomia ``G. Galilei", Universit$\grave{a}$ di Padova, Vicolo dell'Osservatorio 3, 35122 Padova, Italy}
\affil{$^5$ INAF - Osservatorio Astronomico di Brera, via E. Bianchi 46, 23807 Merate (LC), Italy}
\affil{$^6$ Key Laboratory for Research in Galaxies and Cosmology, Shanghai Astronomical Observatory, Chinese Academy of Sciences,80 Nandan Road, Shanghai 200030, China}
\email{liaonh@pmo.ac.cn (NHL); liangyf@pmo.ac.cn (YFL); yzfan@pmo.ac.cn (YZF)}
\begin{abstract}
Narrow line Seyfert 1 galaxies (NLS1s) usually do not host relativistic jets and the detection of their $\gamma$-ray emissions is rarely reported. Evidences for strong relativistic beaming effect have been found in all $\gamma$-ray NLS1s to date and hence their jets are believed to be closely aligned. In this work we perform a systematic analysis of 7-year {\it Fermi} Large Area Telescope (LAT) data of 18 steep radio spectrum NLS1s (S-NLS1s) and report the {\it first} two detections of $\gamma$-ray emissions in this class, increasing the total number of known $\gamma$-ray NLS1s to 9. A strong and stable $\gamma$-ray signal (over 8$\sigma$) is produced by SDSS J144318.56+472556.7, and a significant $\gamma$-ray emission (4.5$\sigma$) of SDSS J130522.74+511640.2 has been detected in a monthly flare in November 2013. A thorough investigation of their multiwavelength properties from radio to X rays has been performed, finding no significant variability. Along with their spectral energy distributions and the radio morphology, these facts strongly suggest that these two sources harbor mis-aligned and possibly underdeveloped relativistic jets, which provides valuable targets to reveal the formation and evolution of relativistic jets under extreme circumstances.
\end{abstract}


\section{INTRODUCTION}
Powered by accretion onto super-massive black holes (SMBHs) with masses up to $\rm \sim 10^{10} M_{\odot}$, active galactic nuclei (AGNs, Urry \& Padovani 1995) are the most persistent energetic objects in the universe. As a minority of these objects, the radio-loud AGNs (RLAGNs, i.e. radio-loudness parameter {\it R}$ > $10, where {\it R} is defined as the ratio between the radio 5 GHz to optical $B$ band luminosity; Kellerman et al. 1989), a substantial fraction (as much as tens of percent) of their accretion energies is extracted into the kinetic and magnetic energies of the relativistic, highly collimated jets. Benefited by observations from Compton $\gamma$-ray Observatory together with the first generation of Cherenkov telescopes, it has been recognized that RLAGNs are capable to generate strong $\gamma$-ray emissions (Hartman et al. 1999; Horan \& Weekes 2004). Especially, blazars dominate the extragalactic $\gamma$-ray sky (e.g. Ackermann et al. 2015). They are characterized by the luminous, rapidly variable, and polarized nonthermal continuum emissions, widely accepted as being produced in the relativistic jets oriented close to the line of sight (Blandford \& Rees 1978; Ulrich et al. 1997). In addition to blazars, $\gamma$-ray emissions from radio galaxies as well as steep spectrum radio quasars (SSRQs), the so-called mis-aligned AGNs (MAGNs), whose jet inclination angle is larger than blazars, have been also detected (e.g. Abdo et al. 2010a). An universal two-bump structure in log$\rm \nu F_{\nu}$-log$\rm \nu$ plot appears in their spectral energy distributions (SEDs). The first bump is (likely) by synchrotron emission of relativistic electrons in magnetic fields, while the second bump, extending to $\gamma$ rays, is usually explained as inverse Compton (IC) scattering of soft photons from either inside (i.e. synchrotron self-Compton, SSC) and/or outside of the jet (i.e. external Compton, EC) by the same population of relativistic electrons (Maraschi et al. 1992; Dermer \& Schlickeiser 1993; Sikora et al. 1994; B{\l}a{\.z}ejowski et al. 2000). 

One of the major findings of {\it Fermi}-LAT is that it reveals radio-loud narrow-line Seyfert 1 galaxies (RLNLS1s) as a new class of $\gamma$-ray AGNs, which provides a compelling evidence of their jetted AGN nature (Abdo et al. 2009a, b). NLS1s are a peculiar subclass of AGNs, with classification criterion of a narrow width of the broad Blamer emission lines with FWHM ($\rm H_{\beta}$) $<$ 2000 km $\rm s^{-1}$, along with strong optical $\rm Fe_{II}$ lines and weak forbidden lines (e.g. Pogge 2000). In consideration of their accretion systems with small black hole masses (between $\rm 10^{5}$ and $\rm 10^{8}$ $\rm M_{\odot}$) and high Eddington ratios ($\gtrsim$ 0.1), NLS1s are believed to be an early stage of AGN activity (e.g. Boroson \& Green 1992; Mathur 2000; Collin \& Kawaguchi 2004). On the other hand, normal radio-loud AGNs, such as blazars and radio galaxies, tend to have higher central BH masses and lower Eddington ratios (Sikora et al. 2007). Only 7\% of all NLS1s are radio-loud (Komossa et al. 2006) while the fractions for normal Seyfert galaxies and quasars are significantly higher, $\sim$ 10\%$-$20\% (Ivezi{\'c} et al. 2002). Based on these differences, RLNLS1s serve as an intriguing population of sources, shedding important new insights on the phenomenon of diverse jet production efficiencies among objects that are very similar in all other aspects, and the formation of radio jets under extreme conditions.

In early studies of RLNLS1s, it has been recognized that RLNLS1s can appear in two flavors (e.g. Zhou et al. 2006; Komossa et 2006; Yuan et al. 2008). Some show a flat radio spectrum (F-NLS1s, radio spectral index $\mid\alpha\mid <$ 0.5, throughout this work we refer to a spectral index $\alpha$ as the energy index such that $S_{\nu}\propto\nu^{-\alpha}$, corresponding to a photon index $\Gamma =\alpha+1$), while others have a steep radio spectrum (S-NLS1s, $\alpha \geq$ 0.5). The formers, in analog of blazars (e.g. Chen \& Bai 2011; Foschini et al. 2015), have a compact radio morphology indicative of an aligned jet. The latters, instead, have an extended morphology that reminiscent of MAGNs. So far, all RLNLS1s with $\gamma$-ray detections are F-NLS1s (Abdo et al. 2009a, b; Foschini et al. 2015; Yao et al. 2015; D'Ammando et al. 2015). In contrast, the role of S-NLS1s is still an open question. Corresponding to the orientation based unified scheme of RLAGNs, S-NLS1s could be the parent population of the F-NLS1s (Foschini 2011; Berton et al. 2015). However, only a handful of RLNLS1s with jets seen at large angles have been found in detailed VLBI studies (e.g. Doi et al. 2012; Gu \& Chen 2010; Richards \& Lister 2015; Gu et al. 2015). This number is far below the expectation number that is calculated as $\rm \sim 2\Gamma^{2}$ times the number of beamed sources ($\gtrsim$ 1000, assuming $\Gamma$ is about 10). Moreover, S-NLS1s could be of particular interest that they offer a different perspective than the F-NLS1s. New $\gamma$-ray phenomenon might emerge since the core jet emission is significantly suppressed due to the mild Doppler beaming effect. Recently, detailed multiwavelength studies on individual S-NLS1 including analyses of their {\it Fermi}-LAT data have been performed, but failed to yield any valid $\gamma$-ray detection (Caccianiga et al. 2014; Komossa et al. 2015). Systematical studies of the radio and optical behaviors of S-NLS1s are also performed (e.g. Berton et al. 2015). However, a systematical analysis of their $\gamma$-ray properties is still lacking.

In this paper, we perform a systematical analysis on the LAT data of 18 S-NLS1s selected in Berton et al. (2015). We report discoveries of $\gamma$-ray emissions from SDSS J130522.74+511640.2 and SDSS J144318.56+472556.7 (hereafter J1305+5116 and J1443+4725, respectively) together with further investigations of their multiwavelength radiation properties. This work is organized as follows: in Section 2 the basic information of the sources and routines of {\it Fermi}-LAT and Swift data analyses are introduced; in Section 3 the results of the $\gamma$-ray data analysis is presented; in Section 4 and 5 the multiwavelength properties and results of SED modeling of the two $\gamma$-ray S-NLS1s are provided, respectively. Finally, discussions and summary are given in Section 6.

\section{THE SAMPLE AND DATA ANALYSES}

\subsection{The sample}
Basic information together with radio and optical properties of the 18 S-NLS1s are summarized in Table 1. All sources exhibit NLS1-like optical spectra and the only exception is J1413-0312 that is identified as a NLS1 by its infrared spectrum (Nagar et al. 2002). Other selection criterions also include a steep radio spectral index $\alpha \geq$ 0.5 and a radio-loudness R $>$ 10. Note that all sources depart significantly from the galactic plane ($\vert b\vert > 25^{\circ}$), for which the {\it Fermi}-LAT data analysis suffers from less uncertainty of the Galactic $\gamma$-ray template. The redshift distribution of the sample is from 0.006 (J1413-0312) to 0.788 (J1305+5116) and about half sources have redshifts lower than 0.1. Only two sources (J1435+3131 and J1443+4725) have radio-loudness above 1000, making them as our primary targets. 
\subsection{LAT Data Analysis}
{\it Fermi}-LAT is a pair-conversion $\gamma$-ray telescope sensitive to photon energies greater than 20 MeV (Atwood et al. 2009). Benefited from its large peak effective area ($\sim$8000 $\rm cm^{2}$ for 1 GeV photons), wide field of view ($\simeq$ 2.4 sr) and relatively high angular resolution (68\% containment radius better than $\rm 1^{\circ}$ at 1 GeV), together with the routine survey mode, LAT performs a complete and uniform coverage of the sky every 3 hours and thus serves as an important tool to detect $\gamma$-ray emission from jetted AGNs.

The \texttt{SOURCE} type (i.e. \texttt{evclass} 128 and \texttt{evtype} 3) \texttt{Pass} 8 data with energy range between 100 MeV and 500 GeV, which lead to an improvement of 30\% - 50\% enhanced differential point-source sensitivity than the former data set (Atwood et al. 2013), were collected from 2008 August 4th to 2015 August 4th. The {\it ScienceTools} software package version \texttt{v10r0p5} together with the instrument response functions of \texttt{P8R2\_SOURCE\_V6} were adopted, as well as the galactic diffuse model \texttt{gll\_iem\_v06.fit} and the isotropic diffuse emission template \texttt{iso\_P8R2\_SOURCE\_V6\_v06.txt}. Then $\gamma$-ray flux and spectrum were extracted by \texttt{unbinned} likelihood algorithm (Mattox et al. 1996) implemented in the \texttt{gtlike} task after filtration performed with \texttt{gtselect} and \texttt{gtmktime} tasks. During the extraction, all sources from the {\it Fermi} LAT third source catalog (3FGL, Acero et al. 2015) within $20^{\circ}$ of the target position were considered. Especially, for sources within $10^{\circ}$ region of interest (ROI), their flux and spectral parameters of sources were set free. On the other hand, those of other sources were fixed at the 3FGL values. The normalization factors of the two diffuse backgrounds were also let to be free.

Firstly, beyond the initial background model, we added an assumed $\gamma$-ray source located at a radio position corresponding to each source in the sample, whose spectral template was set as power law function. After a fit of the 7-year entire data, we checked the Test Statistic (TS) value of the target. If a relatively significant TS value (i.e. $>$ 15) was measured, the localization of this signal was then derived by the \texttt{gtfindsrc} task. A $10^{\circ}\times10^{\circ}$ scale TS map with each pixel size of $0.25^{\circ}$ was obtained to check whether new background sources appeared and a following zoomed-in TS map at scale of $4^{\circ}\times4^{\circ}$  with $0.1^{\circ}$ per pixel was used to investigate the local residue of the fit. Any new significant sources (i.e. TS $>$ 25) were added to the background model with Power-law spectral model and location derived from \texttt{gtfindsrc} task. Then the updated background model was refitted to obtain the final result. After the entire fit, a monthly $\gamma$-ray light curve was extracted for each source. During this temporal analysis, the spectral indexes of weak background sources (TS $<$ 2500) were fixed to the values of the entire fit, while they were removed from the background list when their TS $<$ 1.

\subsection{Swift Data Analysis}
The multiwavelength observatory {\it Swift} carries three instruments (Gehrels et al. 2004), including the Burst Alert Telescope, the X-ray Telescope (XRT), and the UV/Optical Telescope (UVOT). When available, we analyzed both XRT and UVOT data with the FTOOLS software version 6.19. In 2012, the {\it Swift}-XRT visited J1443+4725 11 times with a total exposure time of about 6.4 ks. All XRT data were taken in the photon-counting mode, and were processed with the task of \texttt{xrtpipeline}. We extracted the X-ray photons at the source position within 30\arcsec and just obtained 45 counts which are not sufficient for spectral modeling. In this case, the stacked image was analyzed with the X-ray image analysis program, \texttt{XIMAGE}\footnote{http://www.swift.ac.uk/analysis/xrt/xrtcentroid.php}, and the count rates around the source coordinate were estimated to be of $(7.1\pm1.3) \times 10^{-3}$ and $(5.4\pm1.2) \times 10^{-3}$ cts s$^{-1}$ in 0.2$-$10 keV and 0.2$-$2 keV, respectively. Assuming an absorbed power-law model with the column density fixed to the Galactic value $\rm N_{H}$=$1.47\times10^{20}$ cm$^{-2}$ (see also Yuan et al. 2008), we determined the photon index of $\Gamma \sim 1.9$ and the unabsorbed flux in 0.2$-$10 keV of $\sim$ $2.8\times10^{-13}$ ergs cm$^{-2}$ s$^{-1}$with the help of \texttt{WebPIMMS}\footnote{http://heasarc.gsfc.nasa.gov/cgi-bin/Tools/w3pimms/w3pimms.pl}. For J1305+5116, only 25 photons were collected in all 3 {\it Swift}-XRT observations with a total exposure time of about 4.2 ks. Using the same method described above, we estimated a count rate of $(4.0\pm1.2)\times10^{-3}$ cts s$^{-1}$ and an unabsorbed flux of $\sim$ $1.5\times10^{-13}$ ergs cm$^{-2}$ s$^{-1}$ in 0.2$-$10 keV by using $\Gamma = 1.9$ and $\rm N_{H}$=$ 1.12\times10^{20}$ cm$^{-2}$.

The UVOT has six filters: $V$, $B$, $U$, $UVW1$, $UVM2$ and $UVW2$. When available, extensions were summed within each image using \texttt{uvotimsum}. To determine the source magnitude, we performed aperture photometry for all filters in the sky image using \texttt{uvotsource} with a source extraction region from 5\arcsec, and the background emissions were extracted from a neighboring source-free region. For the UV/optical magnitudes, correction of the Galactic extinction using the E($B$-$V$) values from Schlafly \& Finkbeiner (2011) and the extinction laws from Cardelli et al. (1989) have been performed. For J1305+5116, the averaged extinction corrected magnitudes (in AB system) were $V= 17.34\pm0.12$, $B =17.50\pm0.08$, $U = 17.48\pm0.06$, $UVW1 =17.68\pm0.04$, $UM2 = 17.39\pm0.06$, and $UVW2 =17.59\pm0.05$, respectively. The corresponding magnitudes of J1443+4725 in the six bands were $V= 18.03\pm0.17$, $B =18.02\pm0.10$, $U = 18.07\pm0.05$, $UVW1 =18.27\pm0.06$, $UM2 = 18.27\pm0.07$, and $UVW2 =18.29\pm0.07$, respectively. Magnitudes of single exposures are listed in Table 3.

\section{RESULTS}
Results of the 7-year $\gamma$-ray data analysis are presented in Table 2. For majority of the S-NLS1s in our sample, the corresponding added central $\gamma$-ray sources have TS $<$ 1. Significant $\gamma$-ray excesses with TS values $\gtrsim$ 25 (the corresponding significance is $\simeq 4.1\sigma$ with degree of freedom of 4) around the radio positions of S-NLS1s only hold in two sources, J1305+5116 and J1443+4725. Note that our null result on J1432+3014 is consistent with a previous study that found no significant $\gamma$-ray emission of this source (Caccianiga et al. 2014). For some sources (e.g. J1413-0312 and J2314+2243), initially strong signals can be obtained but following $\gamma$-ray localization analyses and TS maps suggest that the $\gamma$ rays are actually from new nearby background sources. After extracting the new neighbors, TS values of the assumed $\gamma$-ray sources are significantly reduced, suggesting no evidences of strong $\gamma$-ray emissions from the targets. Although J2314+2243 was suggested as a potential $\gamma$-ray emitter (Komossa et al. 2015), no evidence is found here. In temporal analysis, we do not find any significant $\gamma$-ray flare events at short timescale from the monthly light curves except J1305+5116 , see Figure 1, 2 and 3. The highest TS values of the monthly bins are about 10, which is too low to perform further investigations. Detailed analyses of J1305+5116 and J1443+4725 are specially described below. 

\subsection{Detecting $\gamma$-ray emission from J1305+5116}
J1305+5116 was initially detected in radio bands (e.g. Hales et al. 1990; Condon et al. 1998). It was then recognized as a quasar based on its SDSS optical spectrum (Schneider et al. 2005). It was further classified as a NLS1 (broad H$\beta$ emission line width (FWHM) of (1925$\pm$53) km $\rm s^{-1}$, along with $R_{4570}$ $\sim$ 0.65 and [OIII]$\lambda$5007/H$\beta$ $\sim$ 0.25; Zhou et al. 2006; Yuan et al. 2008). Its radio spectral index between 1.4 and 5~GHz is 0.5 and becomes steeper (0.56) between 151~MHz and 4.85~GHz (Yuan et al. 2008; Gu et al. 2015). Its radio loudness is about 200 (Yuan et al. 2008). Interestingly, it is the farthest source in the sample (z = 0.788) as well as one with the largest black hole mass ($\rm \sim3\times 10^{8}$ $\rm M_{\odot}$; Berton et al. 2015).

Initially, a fit of the entire 7-year LAT data provides a relatively high TS value ($\simeq$30) for the assumed $\gamma$-ray source located at the radio position of J1305+5116, see Figure 4. However, the actual location jumps about $0.6^{\circ}$ away after performing a $\gamma$-ray localization analysis, which suggests $\gamma$ rays in this direction is mainly from a new neighbor. Nevertheless, such a single source can not account for the total residual and the remaining part can be reasonably explained by another $\gamma$-ray source spatially coinciding with J1305+5116 (i.e. R.A. $\rm 196.171^{\circ}$ and DEC. $\rm 51.397^{\circ}$, with a 95\% C. L. error radius of $\rm 0.363^{\circ}$). If both these sources are considered in the fit, the TS values are 26.3 and 29.7 for the target and the neighbor, respectively. On the other hand, if the neighbor alone is selected, its TS value reaches up to 50, with a slightly softer spectrum ($\Gamma\simeq2.5$) than the former case ($\Gamma\simeq2.1$). The $\delta_{likelihood}$ between these two fits is about 10, suggesting that the actual TS value of the target is only about 20 rather than the apparent value of 26. Since the LAT PSF for 100~MeV photons is a few degrees but the angular separation is only $0.6^{\circ}$, it is not surprising that these two sources can not be well distinguished. Thus, it is not significant enough to claim a detection of $\gamma$-ray emissions of J1305+5116 based on the results of the entire 7-year LAT data analysis.

Due to the limited number of photon events, focusing higher energy photons that possess better angular resolution is not helpful. Therefore, a temporal analysis is performed to search for any possible flare events at short timescale. Luckily, such a behavior indeed happens in the monthly $\gamma$-ray light curve, see Figure 3. In November 2013 (i.e. {\it Fermi} machine eject time range from 404660413 to 407815247), a relatively strong $\gamma$-ray source with TS value of 28.5 ($\simeq$ 4.5$\sigma$) is emerging. Since the neighbor is relatively faint and it is unlikely that both the target and neighbor are at the flaring state simultaneously, the contamination from the neighbor is negligible at this time. The impact of other background sources was also checked. The only source that is significantly brighter than the central excess is about $9^{\circ}$ away. Together with its high galactic latitude (i.e. $\rm \simeq 65^{\circ}$), the $\gamma$-ray signal is robust at this time. Following $\gamma$-ray localization analysis provides a position of R.A. $\rm 196.787^{\circ}$ and DEC. $\rm 51.196^{\circ}$, with a 95\% C. L. error radius of $\rm 0.332^{\circ}$. The angular separation between the $\gamma$-ray source and the radio position of J1305+5116 is $\rm 0.288^{\circ}$. Radio-loud AGN is supposed to be its ideal counterpart for such a high Galactic latitude $\gamma$-ray source. We also sought other potential counterparts through the SIMBAD database\footnote{http://simbad.u-strasbg.fr/simbad/} but J1305+5116 is the only known radio-loud AGN within the $\gamma$-ray localization error radius. There is a flat radio source NVSS J130448+505624 around but $\rm 0.449^{\circ}$ away from the $\gamma$-ray position, and the separation angle increases to $\rm 0.530^{\circ}$ when the second nearest known RLAGN SDSS J130637.39+514312.0 (a radio galaxy) is considered. Based on these facts, we conclude the detection of $\gamma$-ray emission of J1305+5116.

A single powerlaw function is adopted to describe the $\gamma$-ray spectrum of J1305+5116 during the flaring state:
\begin{equation}
 \frac{dN}{dE}=(1.07\pm0.28)\times10^{-11}(\frac{E}{491.88~{\rm MeV}})^{-(2.39 \pm 0.23)},
\end{equation}
with an integrated flux of $\rm (3.46\pm 1.11)\times10^{-8}$ photons $\rm cm^{-2}$ $\rm s^{-1}$. The corresponding apparent isotropic $\gamma$-ray luminosity is $\rm (6.0\pm 1.8)\times10^{46}$ erg $\rm s^{-1}$, assuming a redshift of 0.788 (in this work we take a $\Lambda$CDM cosmology with $H_{0}=67~{\rm km~ s^{-1}~Mpc^{-1}}$, $\Omega_{\rm m}=0.32$, and $\Omega_{\Lambda}=0.68$; Planck Collaboration et al. 2014). By comparison with the 7-year averaged flux, $\rm (4.21\pm 1.45)\times10^{-9}$ photons $\rm cm^{-2}$ $\rm s^{-1}$ with photon index of $2.66 \pm 0.20$, a variability amplitude of nearly one order of magnitude is suggested. No significant variability of the $\gamma$-ray spectral indexes between different flux states are found due to the relatively large error bars. In addition to the monthly light curve, a further daily $\gamma$-ray light curve corresponding to the flaring state has been extracted, but no hint of intraday $\gamma$-ray variability is obtained.  

\subsection{Detecting $\gamma$-ray emission from J1443+4725}
In analog with J1305+5116, J1443+4725 was discovered by a radio survey (Bologna Survey at 408~MHz; Ficarra et al. 1985). Serving as a relatively bright radio source, it was then included in several other radio catalogs, for example the 6th Cambridge catalog (Hales et al. 1988) and the 87GB catalog (Gregory \& Condon 1991). Its optical counterpart was identified by APM scans (McMahon et al. 2002), and embraced in the fourth edition of the Sloan Digital Sky Survey (SDSS) Quasar Catalog (Schneider et al. 2007). In X rays, it was listed in the ROSAT ALL-SKY Survey Faint Source Catalog (Voges 2000). In consideration of its strong radio emission and typical NLS1 behaviors in its optical spectrum (i.e. broad H$\beta$ emission line width (FWHM) of (1848$\pm$113) km $\rm s^{-1}$, along with $R_{4570}$ $\sim$ 1.5 and [OIII]$\lambda$5007/H$\beta$ $\sim$ 0.3), J1443+4725 was suggested as a member of a population of radio loud NLS1s which likely exhibit strong jets (Yuan et al. 2008). J1443+4725 exhibits a steep radio spectrum ($\rm \alpha_{1.4~GHz - 4.85~GHz} = 0.67$, $\rm \alpha_{151~MHz - 4.85~GHz} = 0.60$; Yuan et al. 2008; Gu et al. 2015). It is very radio-loud ($R$=1331; Yuan et al. 2008), an uncommon feature among NLS1s.

An isolated significant $\gamma$-ray source against the background appears around the radio position of J1443+4725 from the analysis of the total 7-year LAT data (see Figure 5). Its TS value is 70.2, corresponding to a significance over 8$\sigma$. Since it is far away from the galactic plane ($l\simeq60^{\circ}$), the influence from uncertainty of the Galactic diffuse emission should not be worried. However, there is a $\gamma$-ray neighbor that may cause possible contamination. The brightest $\gamma$-ray source within the ROI is 3FGL J1454.5+5124 which is about $4.5^{\circ}$ away, with a photon flux nearly one order of magnitude higher than that of the source. Benefited from the sharp improvement of the Point Spread Function (PSF) of LAT when energy of the photon events increases\footnote{http://www.slac.stanford.edu/exp/glast/groups/canda/lat\_Performance.htm}, an individual analysis for LAT photons with energy above 500~MeV is performed. At this case, the signal is still significant (TS $\simeq$ 45.5). Considering that the 68\% C.L. of LAT PSF for 500 MeV photons is about $1.5^{\circ}$, the central excess can survive from the $\gamma$-ray neighbor. Therefore, the detection of $\gamma$-ray emission is robust. Localization of the central excess gives a $\gamma$-ray position of R.A. $\rm 220.848^{\circ}$ and DEC. $\rm 47.542^{\circ}$, with a 95\% C. L. error radius of $\rm 0.139^{\circ}$. A similar result is obtained when selecting photon events with energy above 500~MeV. Actually, J1443+4725 is found to be the only known radio-loud AGN within the 95\% C.L. $\gamma$-ray radius with angular separation of $\rm 0.111^{\circ}$. Any flat-spectrum radio sources indicative of potential blazar candidates are not found within the 95\% C.L. $\gamma$-ray radius. The second nearest known radio-loud AGN is a radio galaxy SDSS J144246.29+474129.4 whose angular separation from the $\gamma$-ray location is $\rm 0.182^{\circ}$. And the nearest blazar SDSS J144446.10+474257.7 is $\rm 0.290^{\circ}$ away. Motivated by the robust signal and the results of $\gamma$-ray localization, we conclude that the central significant $\gamma$-ray excess is from J1443+4725.

A simple powerlaw function can provide an acceptable description of 7-year averaged $\gamma$-ray spectrum for J1443+4725:
\begin{equation}
 \frac{dN}{dE}=(1.56\pm0.21)\times10^{-12}(\frac{E}{516.73~{\rm MeV}})^{-(2.64\pm0.12)},
\end{equation}
with an integrated flux of $\rm (7.29\pm 1.32)\times10^{-9}$ photons $\rm cm^{-2}$ $\rm s^{-1}$. No significant improvement is found when complex spectral templates (e.g. LogParabola function) are used. We also extract its $\gamma$-ray SED which agrees with the powerlaw description. Considering its redshift as 0.703, the average isotropic apparent $\gamma$-ray luminosity from 100 MeV to 500 GeV is $\rm (6.9\pm0.9)\times10^{45}$ erg $\rm s^{-1}$. In the monthly $\gamma$-ray light curve, no significant variability can be found, though the target appears to be at a relatively high flux state in the second half of 2012 which explains its absence from the 3LAC (Ackermann et al. 2015).

\section{Multiwavelength properties of J1305+5116 and J1443+4725}
Besides the $\gamma$-ray results, let us further probe J1305+5116 and J1443+4725 in multiwavelength view. They are bright at mid-infrared wavelengths and detected by the Wide-field Infrared Survey Explorer (WISE; Wright et al. 2010) with high S/N in all four bands. The infrared magnitudes listed in the AllWISE Source Catalog\footnote{http://wise2.ipac.caltech.edu/docs/release/allwise/} are W1 = 13.264$\pm$0.024, W2 = 11.979$\pm$0.022, W3 = 8.912$\pm$0.026 and W4 = 6.570$\pm$0.066 for J1305+5116, and W1 = 13.426$\pm$0.024, W2 = 12.304$\pm$0.022, W3 = 9.327$\pm$0.034 and W4 = 6.696$\pm$0.054 for the other. They share similar mid-infrared colors in the W2-W3 and W1-W2 diagram, falling into the WISE $\gamma$-ray Strip (WGS, Massaro et al. 2012) where the $\gamma$-ray F-NLS1s already occupy (Foschini et al. 2015). The appearance of the strong jet emissions at mid-infrared wavelengths indicates that such strong jets could also generate significant $\gamma$-ray emissions, consistent with our LAT data analyses. If the WISE and LAT data can represent synchrotron and IC components, respectively, a low Compton dominance (CD) value ($\lesssim$ 1) is derived. Since EC process is likely responsible for the $\gamma$-ray emissions of NLS1s (e.g. Sun et al. 2015), CD could be used to constrain the Doppler factor of the jet. Now we can set a constraint as 
\begin{equation}
\rm CD=L_{EC}/L_{SYN}=U^{\prime}_{ext}\delta^{2}/U_{B}=1, 
\end{equation}
where $\rm U^{\prime}_{ext}$ is the energy density of the external emission at the rest frame, $\rm U_{B}$ is the the energy density of magnetic field and $\delta$ is the Doppler factor. The external emission is assumed from the broad line region (BLR) or the dust torus, adopted with typical energy densities of $\rm 3\times10^{-2}$ and $\rm 3\times10^{-4}$ erg $\rm cm^{-3}$  (Ghisellini et al. 2012), respectively. If we set the typical strengths of magnetic field of 5 Gauss and 1 Gauss corresponding to the BLR or the dust torus model (Sun et al. 2015), respectively, a rough constraint of the Doppler factor can be given as $\lesssim$ 5. 

Another approach to constrain the Doppler factor is by searching any rapid variability events indicative of significant Doppler beaming effect. In radio bands, only a few exposures for these sources are found and no significant radio variability is detected. For example, J1305+5116 has been observed at 1.4~GHz in the NRAO VLA Sky Survey (NVSS) and the Allen Telescope Array Twenty-centimeter Survey (ATA), getting fluxes of 87.4$\pm$2.7 and 92.7$\pm$4.5 mJy, respectively (Condon et al. 1998; Croft et al. 2010). The same fluxes were measured for J1443+4725 as well, with NVSS and ATA fluxes of 165.8$\pm$5.0 and 150.5$\pm$4.6 mJy. For each source, there are roughly 80 visits from WISE, with which the infrared variability can be investigated. The infrared W1 and W2 bands light curves with PSF profile-fit photometric magnitudes are derived from near-Earth objects WISE (NEOWISE) Single-exposure Database (Mainzer et al.2011), see Figure 6. The data selection is following the standard threads\footnote{http://wise2.ipac.caltech.edu/docs/release/neowise/expsup/sec2\_3.html}, including \texttt{qual\_frame>0} and \texttt{ph\_qual=A} (i.e. detection with S/N over 10). We also exclude those data with the S/N marked as ``null" or the reduced $\chi^2$ of the profile fit photometries larger than 2. No significant infrared variability of J1443+4725 and J1305+5116 has been found, consistent with their variability flags in the AllWISE catalog (i.e. both ``1100"). In optical band, these two sources have been monitored by the Catalina Real-time Transient Survey (CRTS, Drake et al. 2009; Djorgovski et al. 2011). The Photometry is transformed from the unfiltered instrumental magnitude to Cousins V by V = $\rm V_{CSS}$ + $\rm 0.31(B-V)^{2}$ + 0.04\footnote{http://nesssi.cacr.caltech.edu/DataRelease/FAQ2.html\#improve} where the optical color is based on the SDSS observation, and averaged to a single value during each observing night. Each optical light curve consists of about 40 data points, ranging from about 53500 to 56500 MJD, see Figure 3. No significant optical variability is found by usage of $\chi^{2}$ test and the ``normalized excess variance" ($\sigma_{NXS}$; Edelson et al. 2002) method. Besides the CRTS light curve, the $V$ band magnitudes between the {\it Swift}/UVOT and SDSS observations are in agreement. No variability is found in the UV bands of {\it Swift}/UVOT, see Table 3. In X rays, it is not possible to perform a temporal analysis using the {\it Swift}/XRT data alone due to the limited exposures. However, since J1443+4725 had also been detected by ROSAT, comparison between these observations has been checked. The $Swift$/XRT flux is scaled to energy range of 0.1-2.4 keV, $\sim 2\times10^{-13}$ erg $\rm cm^{-2}$ $\rm s^{-1}$, which is comparable with the ROSAT flux, $2.5\times10^{-13}$ erg $\rm cm^{-2}$ $\rm s^{-1}$ (Yuan et al. 2008). After a thorough investigation of the multiwavelength data, no evidence of significant variability is found from radio to X rays in both sources.

In radio, J1305+5116 exhibits a core-jet morphology in the VLBA radio map, including a flat spectral component in the north and a steep spectral jet in the opposite (Petrov 2013; Gu et al. 2015).  The core flux is significantly brighter than the jet flux. At 5~GHz core flux is 15.0~mJy and the rest is 8.9 mJy. The core brightness temperature is about 5.8 $\rm \times 10^{10}$ K at 5~GHz and about 1.6 $\rm \times 10^{11}$ K at 2.3 and 8.4 ~GHz (Petrov 2013; Gu et al. 2015). On the other hand, its FIRST structure is compact (i.e. 1.34 arcsec), suggesting that its linear size (LS) is less than 10 kpc. Similar with J1305+5116, J1443+4725 is also unresolved in the FIRST image, indicative of an upper limit LS of 8.4 kpc. And the total flux density at 5 GHz from the VLBA image is close to that in the single dish GB6 catalog, suggesting that it is very compact in the sub-arcsec scale if the radio flux is not significantly variable (Gu et al. 2015). On mas scale, seven components along the south-west direction are resolved with diffuse emission extending up to 30~mas ($\sim$ 200 pc). None of these components exhibit flat spectra (between 5 and 6.7~GHz), indicating that the core may be hidden in the brightest component and hence its core dominance is rather low ($R_{c}\lesssim$ 0.67, Gu et al. 2015). The brightness temperature at 5~GHz is 2.0 $\rm \times 10^{10}$ K for the brightest jet feature.

\section{SED Modeling}
Since no significant variability displays in the multiwavelength data, except the monthly $\gamma$-ray variability of J1305+5116, their jet emissions are calculated based on the non-simultaneous SEDs, see Figure 7. For J1305+5116, only its 7-year averaged $\gamma$-ray spectrum is included in the modeling since there is no simultaneous complementary data corresponding to the monthly $\gamma$-ray flare. The broadband non-thermal electromagnetic emission is assumed from a compact homogeneous spherical blob with relativistic speed embedded in the magnetic field, where the emitting electrons follow a broken power-law spectrum distribution,
\begin{equation}
N(\gamma )=\left\{ \begin{array}{ll}
                    K\gamma ^{-p_1}  &  \mbox{ $\gamma_{\rm min}\leq \gamma \leq \gamma_{br}$} \\
            K\gamma _{\rm br}^{p_2-p_1} \gamma ^{-p_2}  &  \mbox{ $\gamma _{\rm br}<\gamma\leq\gamma_{\rm max}$,}
           \end{array}
       \right.
\end{equation}
The parameters of this model include information of electrons, the normalization of the particle number density $K$, electron break energy $\gamma_{br}$, the minimum and maximum energies, $\gamma_{min}$ and $\gamma_{max}$, and the indices $p_{1,2}$ of the broken power-law distribution, as well as the magnetic field strength B and the radius of the blob $R_{b}$ that is constrained by the minimum variability timescale, $R_{b} \leq ct_{var}\delta(1+z)^{-1}$. The radius of J1305+5116 is set corresponding to the timescale (i.e. 30 days) of the monthly $\gamma$-ray variability. The same timescale is set for J1443+4725 due to lack of its variability information. The leptonic radiation scenario takes into account the synchrotron and IC processes (both the SSC and EC). The infrared emission from the dust torus is assumed to be the main contribution of external photons for the EC process. Its peak frequency is set at 1~eV and energy density $\rm U^{\prime}_{dust}$ is set as $\rm 3\times10^{-4}$ erg $\rm cm^{-3}$ (Ghisellini et al. 2012). The minimum energy of the radiation electron $\gamma_{min}$ is set as 100. In our calculations, we properly account for the synchrotron self-absorption and the Klein-Nishina effect in the IC scattering. $\chi^{2}$-minimization method is used to derive the optimum description. A similar SED modeling strategy has been adopted in previous studies (e.g. Zhang et al. 2012; Liao et al. 2014,2016).

The simple single-zone homogeneous leptonic radiation model provides acceptable descriptions of the SEDs, although the multiwavelength data coverage is sparse, see Figure 7. And the input parameters are listed in Table 4. The radio data are only treated as upper limits in the modeling because they are probably originated in an extended region, rather than the compact region where the high energy photons come from. Since the accretion disk emission becomes dominant at optical/UV wavelengths, the SDSS and {\it Swift}/UVOT data are not considered. Due to the limited number of X-ray photons, their X-ray spectra are poorly constrained and hence the origin of X rays is unknown. Thus here we assume the jet emissions mainly contribute the X rays. It is not surprised that similar parameters are derived for these two sources that share similar SED shapes. Note that the values of the Doppler factor are about 5, consistent with our previous simple estimated from the low Compton dominance value. The jet powers of J1305+5116 and J1443+4725 are also estimated, $\rm 1\times10^{46}$ erg $\rm s^{-1}$ and $\rm 6\times10^{45}$ erg $\rm s^{-1}$, respectively, based on the assumption that one proton corresponds to one relativistic emitting electron and that protons are `cold' in the comoving frame (Celotti \& Ghisellini 2008).

\section{DISCUSSION}
Compared with blazars that the number of their $\gamma$-ray population is up to over 1500, only five sources marked as NLS1s are listed in 3LAC (Ackermann et al. 2015), which becomes a major obstacle for understanding the jet physics of NLS1s. Recently, a $\gamma$-ray detection of FBQS J1644+2619 has been reported by revising the $\gamma$-ray localization (D'Ammando et et al. 2015). Another source, SDSS J122222.55+041315.7, its NLS1-like optical emission line properties has been revealed and thus identified as a new $\gamma$-ray NLS1 (Yao et al. 2015). Since its redshift is 0.966, it is the known farthest $\gamma$-ray NLS1. Our systematical analysis of LAT data of 18 S-NLS1s brings two more $\gamma$-ray NLS1s and makes the total number increase to nine. We also put these two sources into the diagram between $\gamma$-ray luminosity and $\gamma$-ray spectral index and they stay where the known $\gamma$-ray NLS1s occupied, see Figure 8, suggesting that J1443+4725 and J1305+5116 exhibit similar $\gamma$-ray properties with other $\gamma$-ray NLS1s.

However, unlike J1305+5116 and J1443+4725 that are S-NLS1s, other $\gamma$-ray NLS1s exhibit a flat spectral radio index (between 1.4~GHz and 5~GHz, Yuan et al. 2008; Foschini et al. 2015). Moreover, evidences of strong Doppler beaming effect have been found for these F-NLS1s. Firstly, a detection of jet component with superluminal motion (i.e. $\rm \beta_{app}$ = 8.2$\pm$1.5) of SBS 0846+513 provides a direct support that its jet is well aligned (D'Ammando et al. 2012). Another strong evidence is that fast variability through multi-wavelengths indicative of a very compact radiation region has been detected for several $\gamma$-ray F-NLS1s. $\gamma$-ray variability with doubling time of $\sim$ 3~hr has been reported for 1H 0323+342 (Paliya et al. 2014). Intranight optical flux variability of 0.3-0.5 mag as well as fast optical polarization variability of SBS 0846+513 and PMN J0948+0022 are detected (e.g. Liu et al. 2010; Maune et al. 2014). Similar nightly variability has also been found in these two sources at infrared wavelengths (Jiang et al. 2012). In addition to the superluminal jet motion and rapid variability, such blazar-like features also emerge in their radio morphology. All $\gamma$-ray F-NLS1s exhibit core-dominated morphology (e.g. Foschini et al. 2015), see Figure 9, along with high brightness temperatures, up to $\rm \sim10^{14}$ K for PKS 2004$-$447 (e.g. Gallo et al. 2006, but also see Schulz et al. 2016\footnote{PKS 2004$-$447 is included in the F-NLS1s catalog presented by Foschini et al. (2015), but a recent study suggests a steep radio spectrum above 5~GHz along with core-dominated radio morphology (Schulz et al. 2016). Actually, there are some doubt about the classification of this source as a NLS1 since its $\rm Fe_{II}$ emission is significantly weaker than the typical value of NLS1s.}). Last but not least, $\gamma$-ray F-NLS1s share similar SED shapes with FSRQs, showing a high CD value (typically, $\gtrsim$10) even at the low flux state, and the radiation model adopted for FSRQs (i.e. $\delta\sim 10$) provides an well description for $\gamma$-ray F-NLS1s (e.g. Abdo et al. 2009b). In contrast, none of these signs of strong Doppler beaming effect are found for J1305+5116 and J1443+4725. No significant variability from radio wavelength to X rays are seen for these two S-NLS1s. In $\gamma$ rays, any flare events are not detected in monthly light curve of J1443+4725. Though a strong monthly $\gamma$-ray flare indeed appears for J1305+5116, no extreme events of rapid variability at timescale of hours are found. The brightness temperature of the two S-NLS1s is relatively low, $\rm \sim10^{11}$ K, well below the inverse-Compton limit of $\rm 10^{12}$ K (Kellermann \& Pauliny-Toth 1969). These two S-NLS1s share similar SED shape with CD $\lesssim$ 1, indicative of $\rm \delta\lesssim5$ confirmed by the SED modeling. Moreover, the radio flux contribution of the core component of J1443+4725 is below 40\% of the total. Note that J1443+4725 is the only known lobe-dominated $\gamma$-ray NLS1s. This is a strong indication of a mis-aligned jet and thus the core flux is significantly suppressed due to the mild relativistic condition, see Figire 9.

The nearest radio galaxy Centaurus A is the first source recognized as a $\gamma$-ray emitter with a large jet inclination angle (the angle of the sub-parsec jet to our line of sight to be $\rm \sim 50^{\circ}$$\rm -80^{\circ}$; Tingay et al. 1998) by EGRET (Sreekumar et al. 1999). Its $\gamma$-ray emission is confirmed soon after the beginning of operation of {\it Fermi}-LAT (Abdo et al. 2010b,c). Besides the nearby radio galaxies, $\gamma$-ray emissions from SSRQs with relatively high redshifts have also detected by {\it Fermi}-LAT. For example, an apparent isotropic $\gamma$-ray luminosity of $\rm \sim 10^{46}$ erg $\rm s^{-1}$ is reported from a strongly lobe-dominated SSRQ 3C 275.1 in which the core only contributes 10\% of the total radio flux (Liao et al. 2015). In the evolutionary scenario of RLAGN, the age of the source is tightly connected with its size. GigaHertz-Peaked Spectrum (GPS) radio sources with linear size (LS) of $\lesssim$ 1~kpc evolve into Compact Steep Spectrum (CSS) sources with LS $\lesssim$ 10$-$15~kpc, then finally become large-scale ($\gtrsim$20 kpc) mature radio sources (see O'Dea 1998 for a review). Typical turnover frequencies of these young radio sources in their radio spectra are between 0.1-1~GHz. Since their jet lengths are confined possibly within the host galaxies, young radio sources are suggested to be strong $\gamma$-ray emitters because of the dense material and radiation environment and thus can be important probe for the jet-galaxy interaction (Bai 2005; Stawarz et al. 2008; Migliori et al. 2014). One possible candidate is 4C 55.17, that shows both a symmetric radio morphology at kpc scale and a stable $\gamma$-ray emission, but both the blazar and CSS radiation model can explain its SED that leaves the debate on its nature still open (McConville et al. 2011). Recently, the detection of $\gamma$-ray emission from PKS 1718$-$649 is claimed and thus it is suggested as a $\gamma$-ray CSS (Migliori et al. 2016). Since NLS1s served as young sources whose radio lobes might be well developed yet, they could strictly connect to the young radio sources (e.g. Oshlack et al. 2001; Komossa et al. 2006; Gu et al. 2016). And in this case the extended radio emission from S-NLS1s could be difficult for detection, which might explain the lack of known S-NLS1s. This scenario is supported by systematical analyses of two statistically complete samples of CSS and F-NLS1 by comparison of their black hole mass and Eddington ratio distributions (Berton et al. 2016). Detailed analyses of individual sources also confirm the tight link between CSSs and S-NLS1s (e.g. Caccianiga et al. 2014, 2017; Gu et al. 2015, 2016). For the two $\gamma$-ray S-NLS1s here, since their radio morphology is compact (LS $<$ 10~kpc), they could be CSS-like. Since J1443+4725 is not detected in the Very Large Array Low-frequency Sky Survey Redux catalog (VLSSr, Lane et al. 2014), considering the VLSSr 74 MHz 5$\sigma$ unpper limit ($\sim 0.5$ Jy), a turnover around 100~MHz in its radio spectrum might present, supporting its CSS nature (also see Gu et al. 2015).

Doppler beaming effect plays an important role in the generation of $\gamma$-ray emission of AGNs, especially for blazars. On the other hand, MAGNs tend to have milder $\gamma$-ray variability at timescale of months rather than days as well as slower apparent knot speeds than blazars, consistent with their observed large jet inclination angles. So it is not surprising that $\gamma$-ray blazars overwhelmingly outnumber $\gamma$-ray MAGN by roughly 100 to 1 (e.g. Ackermann et al. 2015). For nearby Fanaroff$-$Riley type (FR, Fanaroff \& Riley 1974) I radio galaxies, their $\gamma$-ray emission can be well explained by a gradient Lorentz radiation model corresponding to the observed structured jets (Georganopoulos \& Kazanas 2003; Ghisellini et al. 2005; Abdo et al. 2010c). Interestingly, a significant contribution of $\gamma$-ray emission from extended radio lobes of the nearest radio galaxies has been revealed (Abdo et al. 2010b; Ackermann et al. 2016). For SSRQs, since generally they share similar $\gamma$-ray luminosities with FSRQs, core-jet radiation model with slightly mild relativistic condition (e.g. jet inclination angle $\rm \sim 10^{\circ}$) could work, though the contribution from the hotspot/lobe can not be ruled out (e.g. Liao et al. 2015). For the two $\gamma$-ray S-NLS1s here, another possible radiation scenario is the one proposed for young radio sources in consideration of the tight connection between them and CSS sources. Since a monthly $\gamma$-ray flare is found for J1305+5116 indicative of a relatively compact radiation region, the young radio source and the lobe scenarios are likely impossible. However, all the three radiation scenarios can not be distinguished for J1443+4725 due to lack of any variability information. It is worth noting that 2 $\gamma$-ray S-NLS1s have been detected while the number of $\gamma$-ray F-NLS1s is only 7. On the other hand, the number of $\gamma$-ray blazars is about a hundred times of that of radio galaxies and SSRQs, indicative of a significant number ratio discrepancy of $\gamma$-ray aligned and mis-aligned populations between RLNLS1s and `normal' RLAGNs. If the only difference between these two groups is the down-scaling of the black hole mass, such a discrepancy is unexpected. Since the luminosity amplification of the Doppler beaming effect depends both on the viewing angle and the jet speed, one possible explanation of this discrepancy is that the jet speed of RLNLS1s is generally lower than that of the `normal' RLAGNs. Interestingly, unlike the nearest radio galaxy Centaurus A whose $\gamma$-ray emission is firstly detected in its kind, J1305+5116 and J1443+4725 are the farthest two sources in the sample. It could be a selection effect because J1305+5116 harbors the heaviest black hole in the sample, as well as the highest bolometric luminosity (Berton et al. 2015). For J1443+4725, its black hole mass is relatively high (i.e. $\rm \sim 10^{7.4} M_{\odot}$) and more importantly it is one of the two sources in the sample sharing powerful radio jets ($R >$ 1000). Actually, other S-NLS1s could be detected by {\it Fermi}-LAT in the near future, especially for those have low redshifts (e.g. J1413-0312, or NGC 5506). They could be valuable targets providing close looks of jet energy dissipation and connection between jet and the accretion system under the extreme circumstance.

In summary, we perform a systematical analysis of 7-yr LAT data of 18 S-NLS1s. Although no significant $\gamma$-ray emission is found for majority of the sources, we report the discoveries of $\gamma$-ray emissions from J1305+5116 and J1443+4725, increasing the total number of known $\gamma$-ray NLS1s to 9. The latter is likely stable while the former is not. Further investigation of their multiwavelength variability has been performed, but no significant variability is found from radio to X rays. Moreover, the CD of their SEDs is low, indicative of $\delta\lesssim 5$. Since the contribution of radio core of J1443+4725 is less than 40\% of the total, it is the unique lobe-dominated $\gamma$-ray NLS1s so far. Along with their compact radio morphology, J1305+5116 and J1443+4725 are the first examples of mis-aligned $\gamma$-ray NLS1s and might have a tight connect with the young radio sources, serving as important probes to investigate the formation and evolution of relativistic jets in AGNs.

\acknowledgements
This research has made use of data obtained from the High Energy Astrophysics Science Archive Research Center (HEASARC), provided by $\rm NASA^{\prime}$s Goddard Space Flight Center. This research has also made use of the NASA/IPAC Extragalactic Database and the NASA/IPAC Infrared Science Archive which are operated by the Jet Propulsion Laboratory, California Institute of Technology, under contract with the National Aeronautics and Space Administration. This research makes use of the SIMBAD database, operated at CDS, Strasbourg, France. The CSS survey is funded by the National Aeronautics and Space Administration under grant no. NNG05GF22G issued through the Science Mission Directorate Near-Earth Objects Observations Program. The CRTS survey is supported by the US National Science Foundation under grants AST-0909182 and AST-1313422. This work is supported in part by 973 Program of China under grant 2013CB837000, National Natural Science of China under grants 11525313 (the National Natural Fund for Distinguished Young Scholars), 11433009, 11673013, 11473054 and U1531245, as well as the Science and Technology Commission of Shanghai Municipality (grant 14ZR1447100). NHL thanks Wei-ming Yuan, Ting-gui Wang, Luigi Foschini and Liang Chen for their instructive advice.

\begin{figure}
\centering
\includegraphics[scale=0.7]{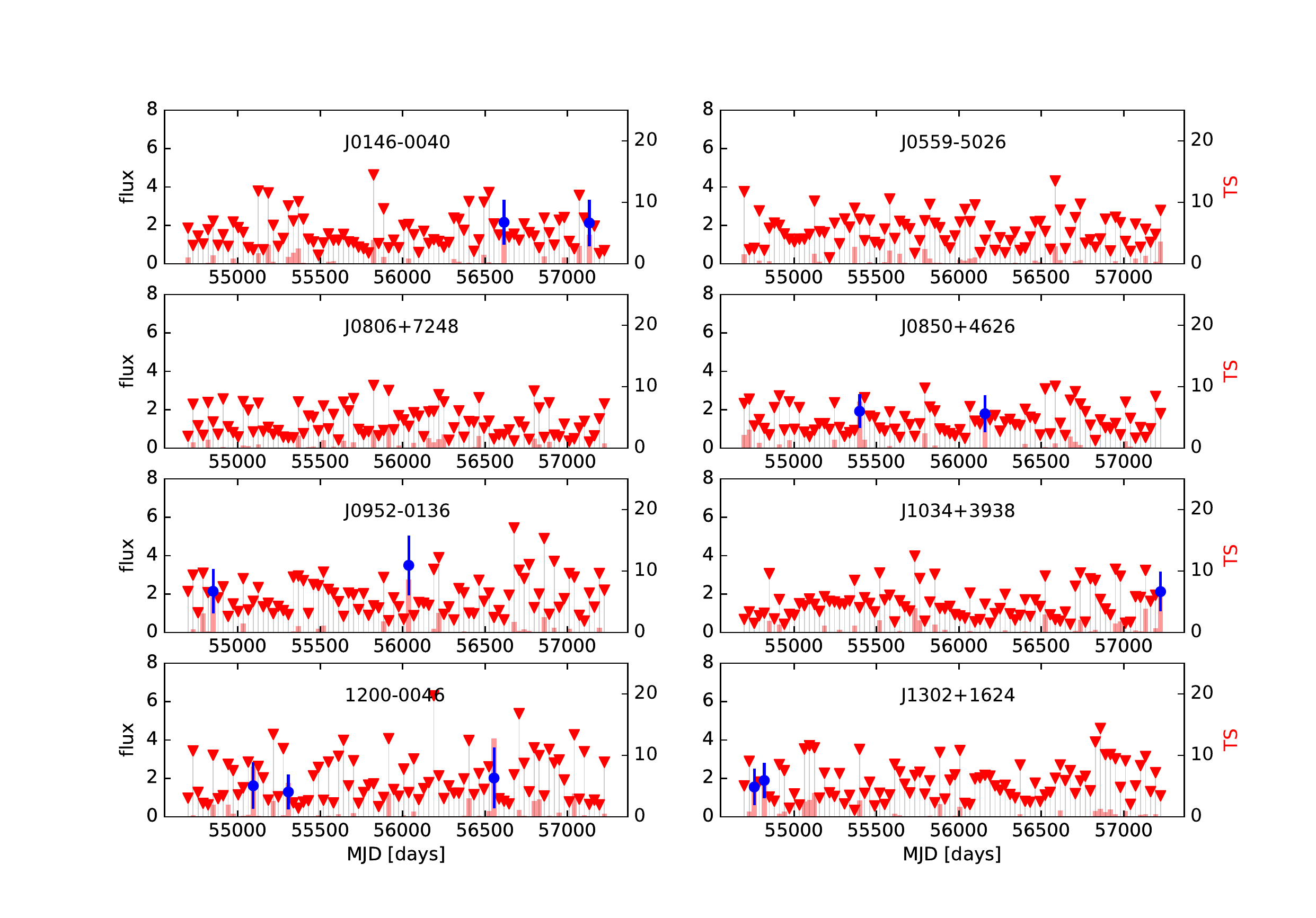}
\caption{Monthly $\gamma$-ray light curves of the S-NLS1s. Blue points and red triangles are fluxes and 2$\sigma$ upper limits in unit of $\rm 10^{-8}$ ph $\rm cm^{-2}$ $\rm s^{-1}$, respectively. TS values are also provided, marked as red bars.}
\label{Fig.1}
\end{figure}

\begin{figure}
\centering
\includegraphics[scale=0.7]{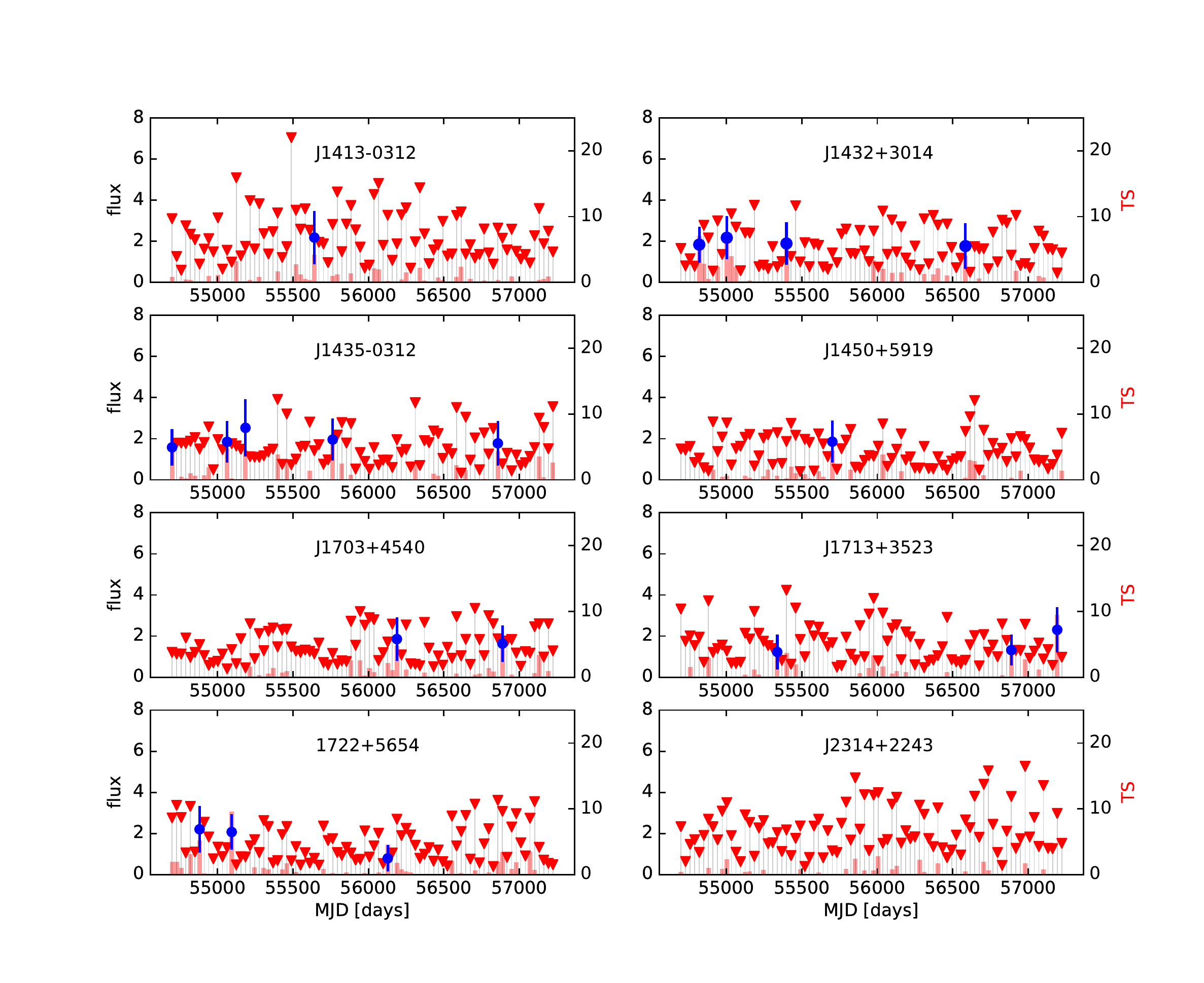}
\caption{Same as in Figure 1.}
\label{Fig.1}
\end{figure}

\begin{figure}
\centering
\includegraphics[scale=0.6]{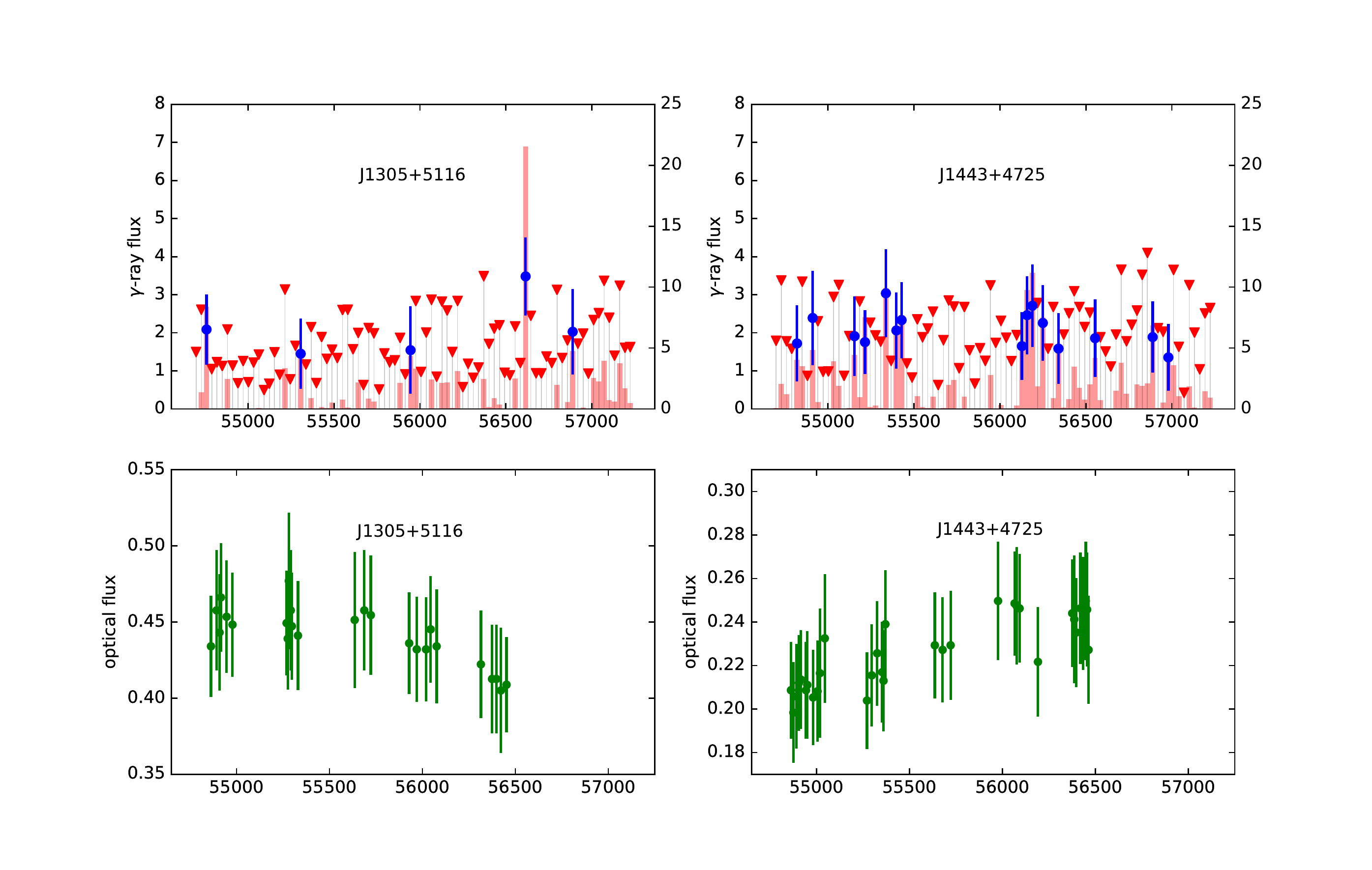}
\caption{$\gamma$-ray and optical light curves of J1305+5116 and J1443+4725. The fluxes, upper limits (in unit of $\rm 10^{-8}$ ph $\rm cm^{-2}$ $\rm s^{-1}$) and the TS values of monthly $\gamma$-ray bin are shown as Blue points, red triangles and red bars, respectively. The extinction corrected nightly $V$ band CRTS fluxes in unit of mJy are marked as green circles.}
\label{Fig.2}
\end{figure}

\begin{figure}
\centering
\includegraphics[scale=0.4]{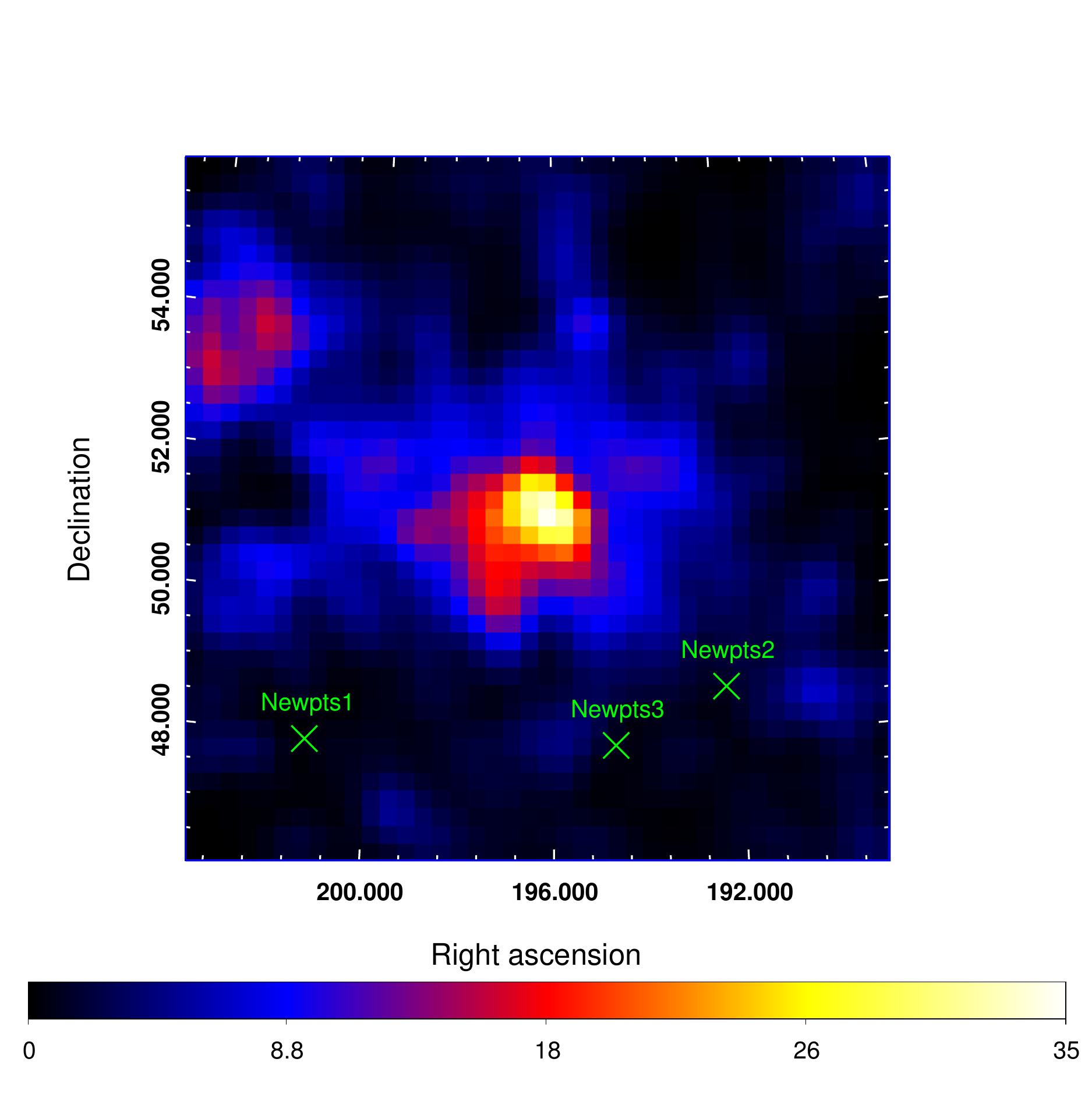}
\includegraphics[scale=0.411]{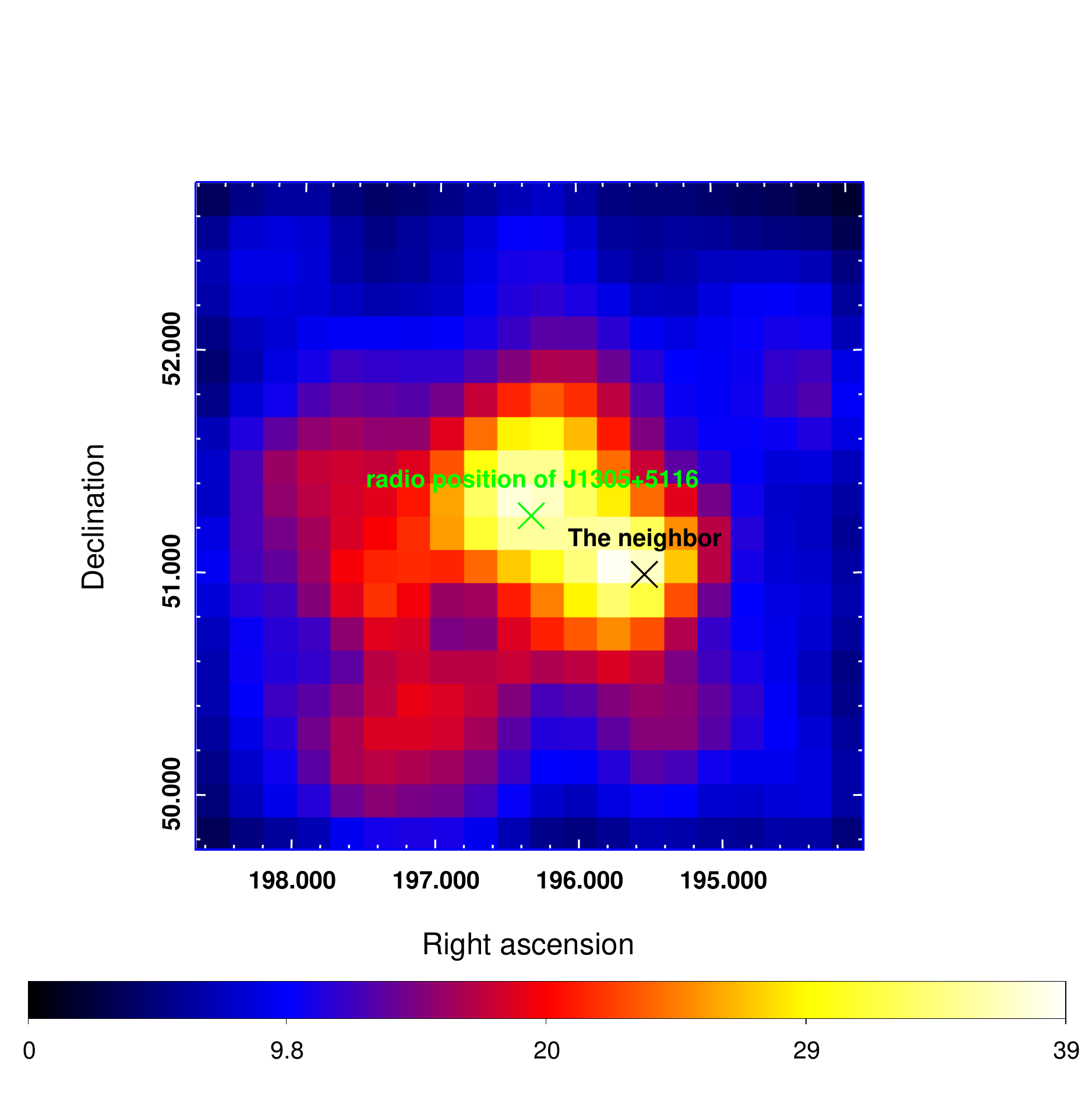}
\includegraphics[scale=0.41]{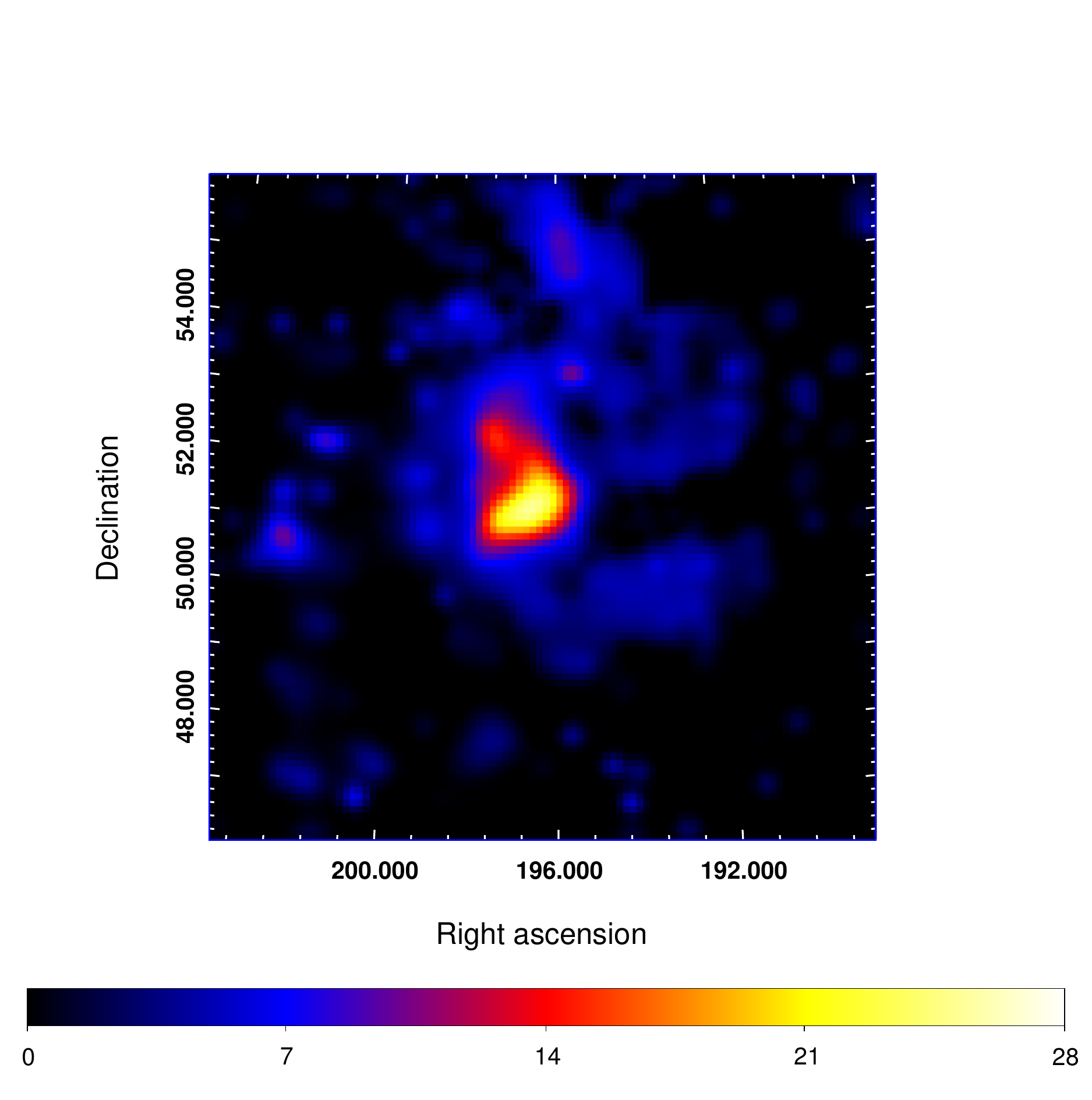}
\includegraphics[scale=0.411]{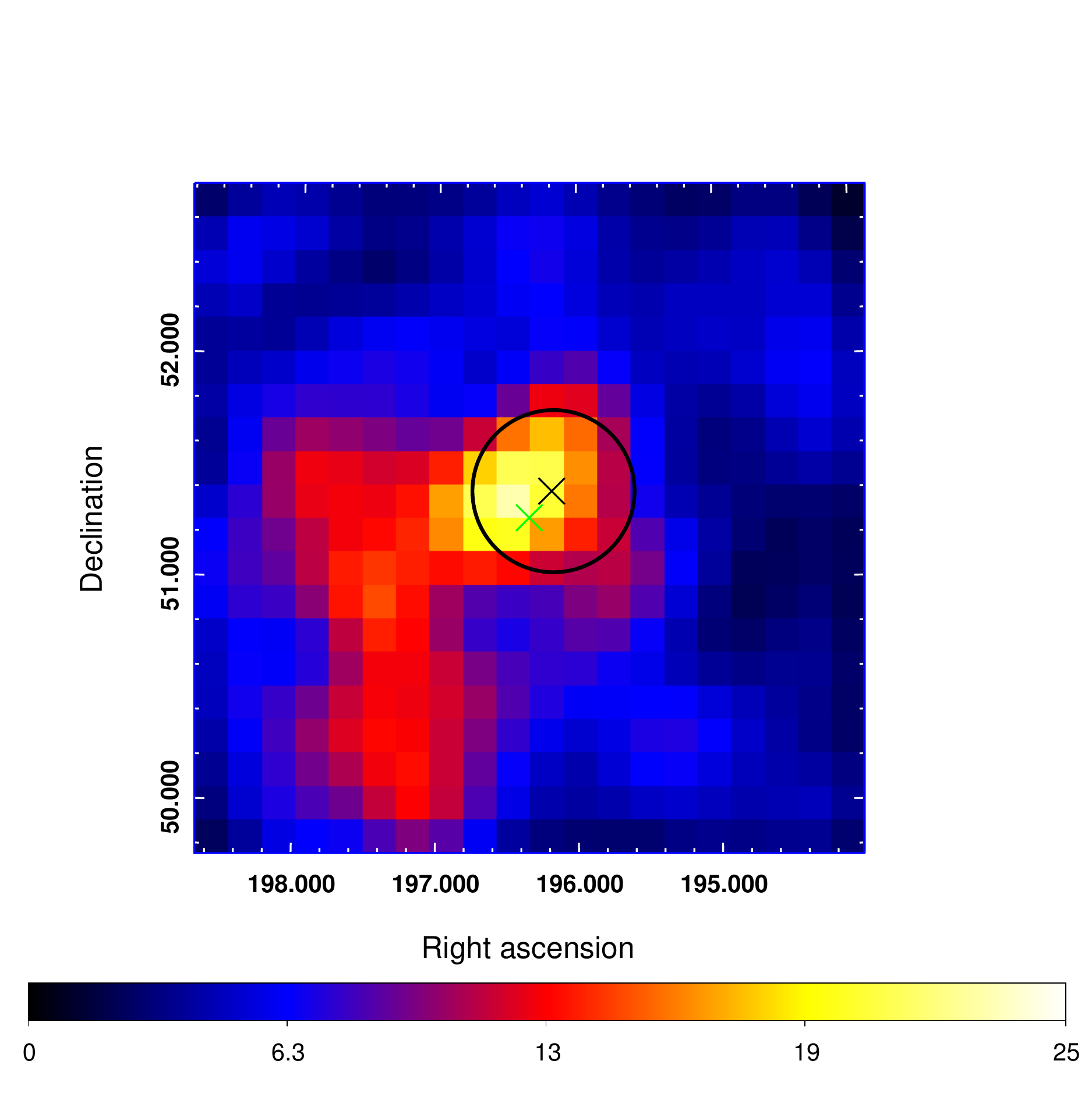}
\caption{TS maps from 100 MeV to 500 GeV centered at J1305+5116. The left panels are TS maps in scale of $10^{\circ} \times 10^{\circ}$ region ($0.25^{\circ}$ per pixel) with diffuse backgrounds as well as both 3FGL and additional sources subtracted, corresponding to analyses of 7-yr LAT data (upper) and the monthly $\gamma$-ray flare (bottom). The upper right panel is a zoomed-in frame of the upper left panel in scale of $4^{\circ} \times 4^{\circ}$ region ($0.1^{\circ}$ per pixel), where $\gamma$-ray position of the close neighbor (black) and the radio position of J1305+5116 (green) are marked. The bottom right panel is as same as the upper right panel but with the close neighbor subtracted. The black x label and the circle are the $\gamma$-ray location of the rest residual and its 95\% C.L. error radius, respectively. The green x label is the radio position of J1305+5116. All maps are smoothed with $\sigma$=$0.2^{\circ}$ Gaussian function.}
\label{Fig.2}
\end{figure}

\begin{figure}
\centering
\includegraphics[scale=0.6]{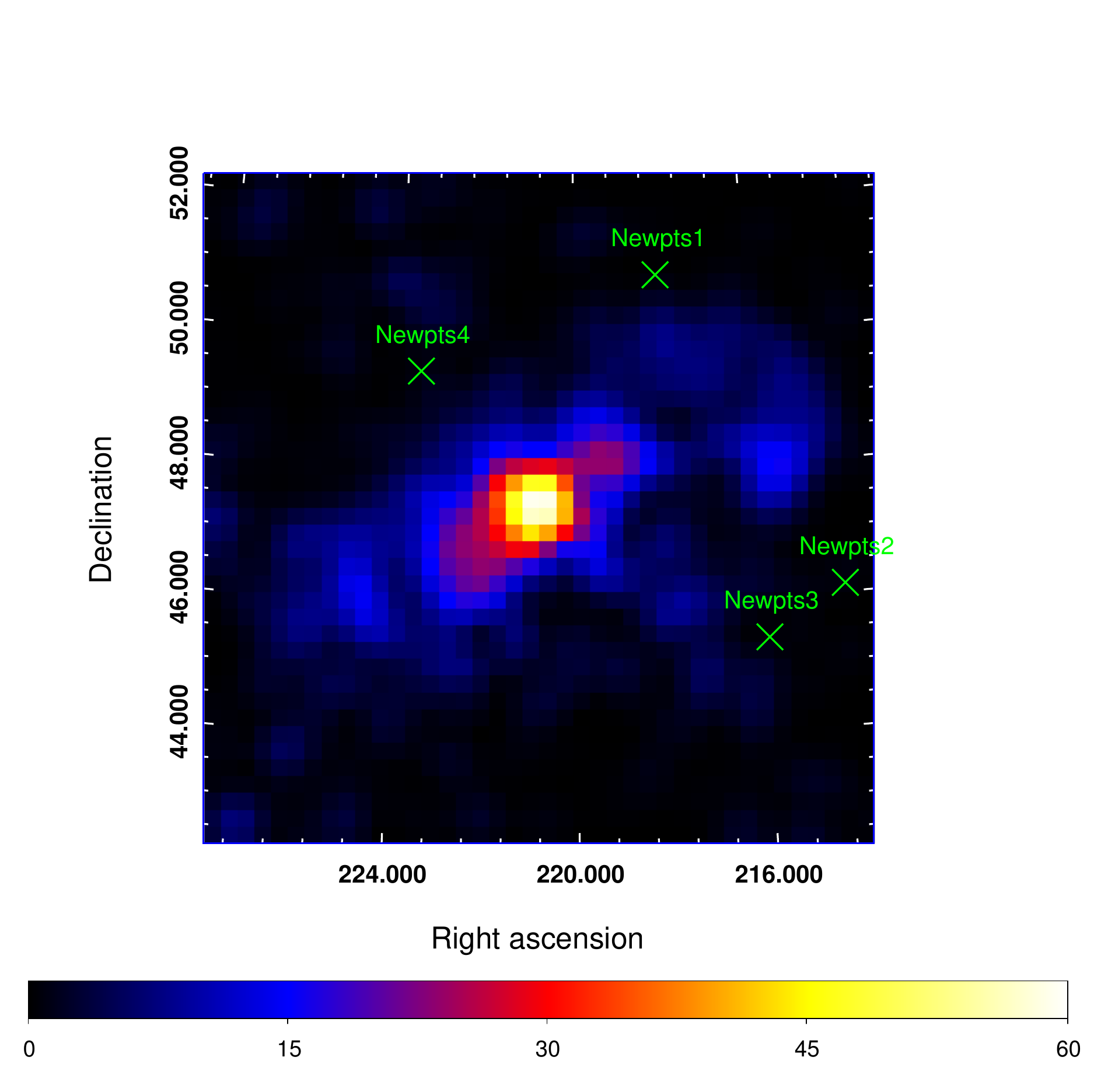}
\caption{TS map from 100 MeV to 500 GeV for $10^{\circ} \times 10^{\circ}$ region ($0.25^{\circ}$ per pixel) centered at J1443+4725 . It is for the model with diffuse backgrounds and both 3FGL and additional sources subtracted. The map is smoothed with $\sigma$=$0.2^{\circ}$ Gaussian function.}
\label{Fig.1}
\end{figure}

\begin{figure}
\centering
\includegraphics[scale=0.7]{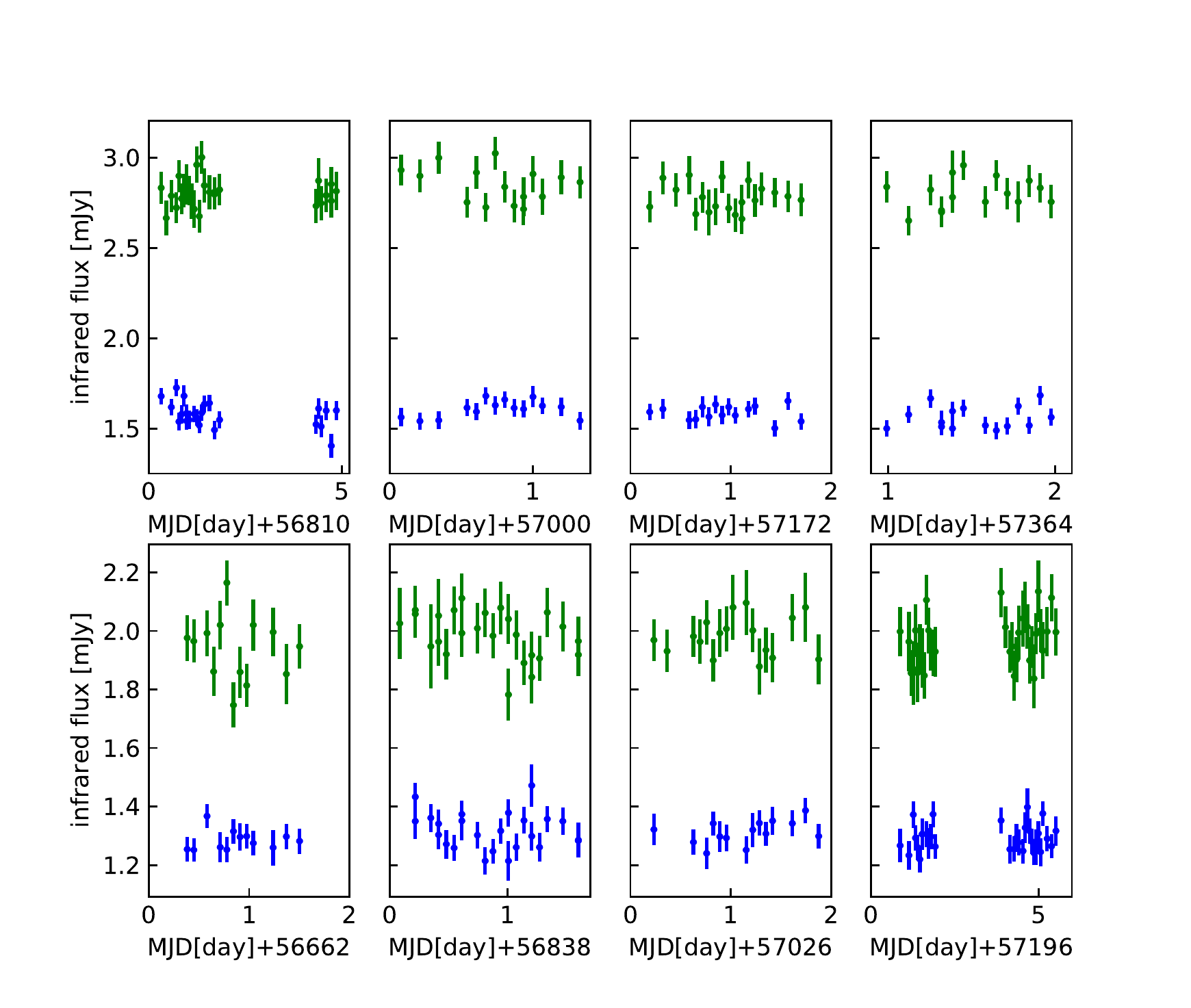}
\caption{{\it WISE} infared light curves of J1305+5116 (upper panel) and J1443+4725 (bottom panel). The blue and green points correspond to fluxes of the $W1$ and $W2$ bands, respectively.}
\label{Fig.1}
\end{figure}

\begin{figure}
\centering
\includegraphics[scale=0.7]{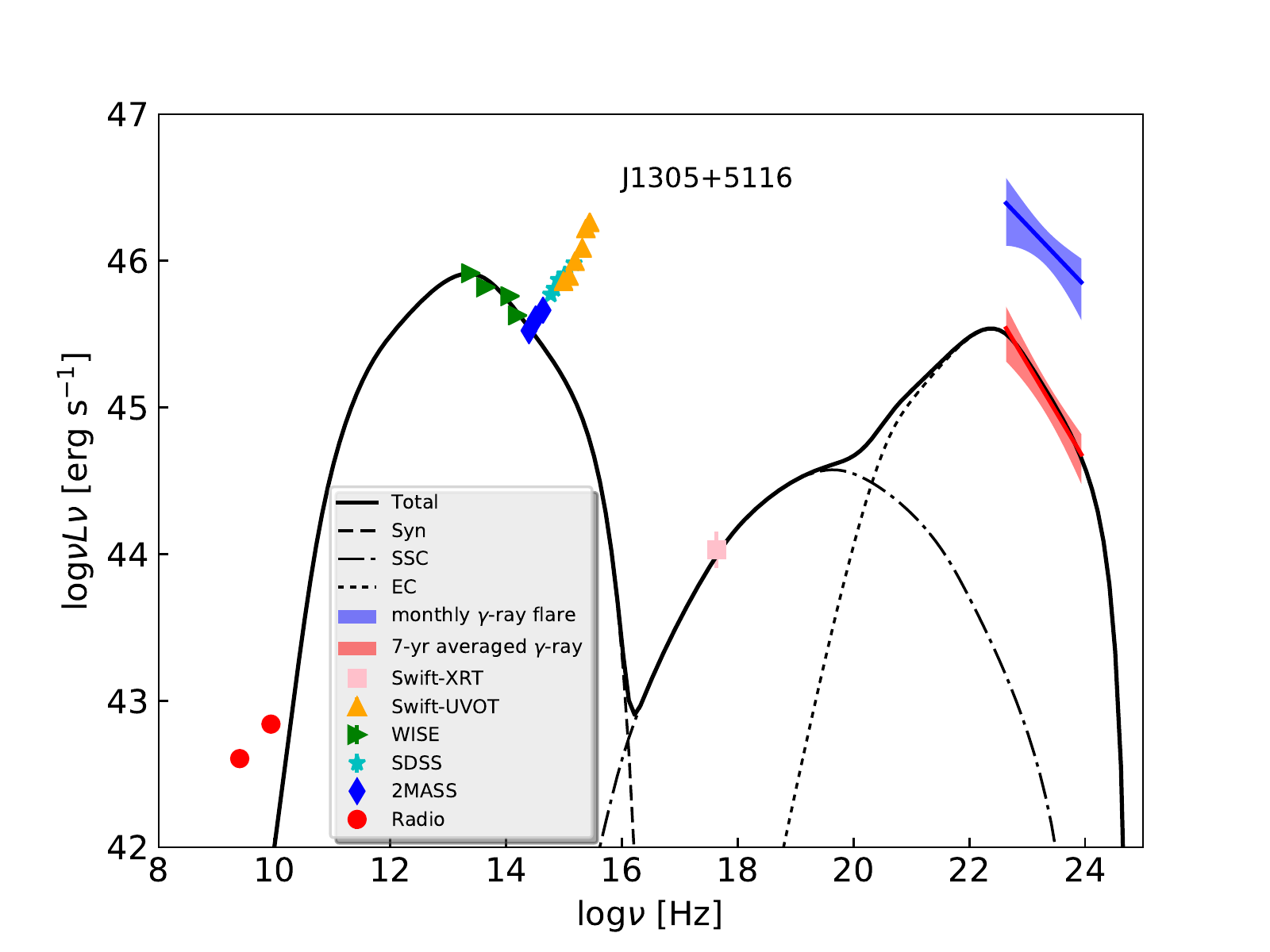}
\includegraphics[scale=0.7]{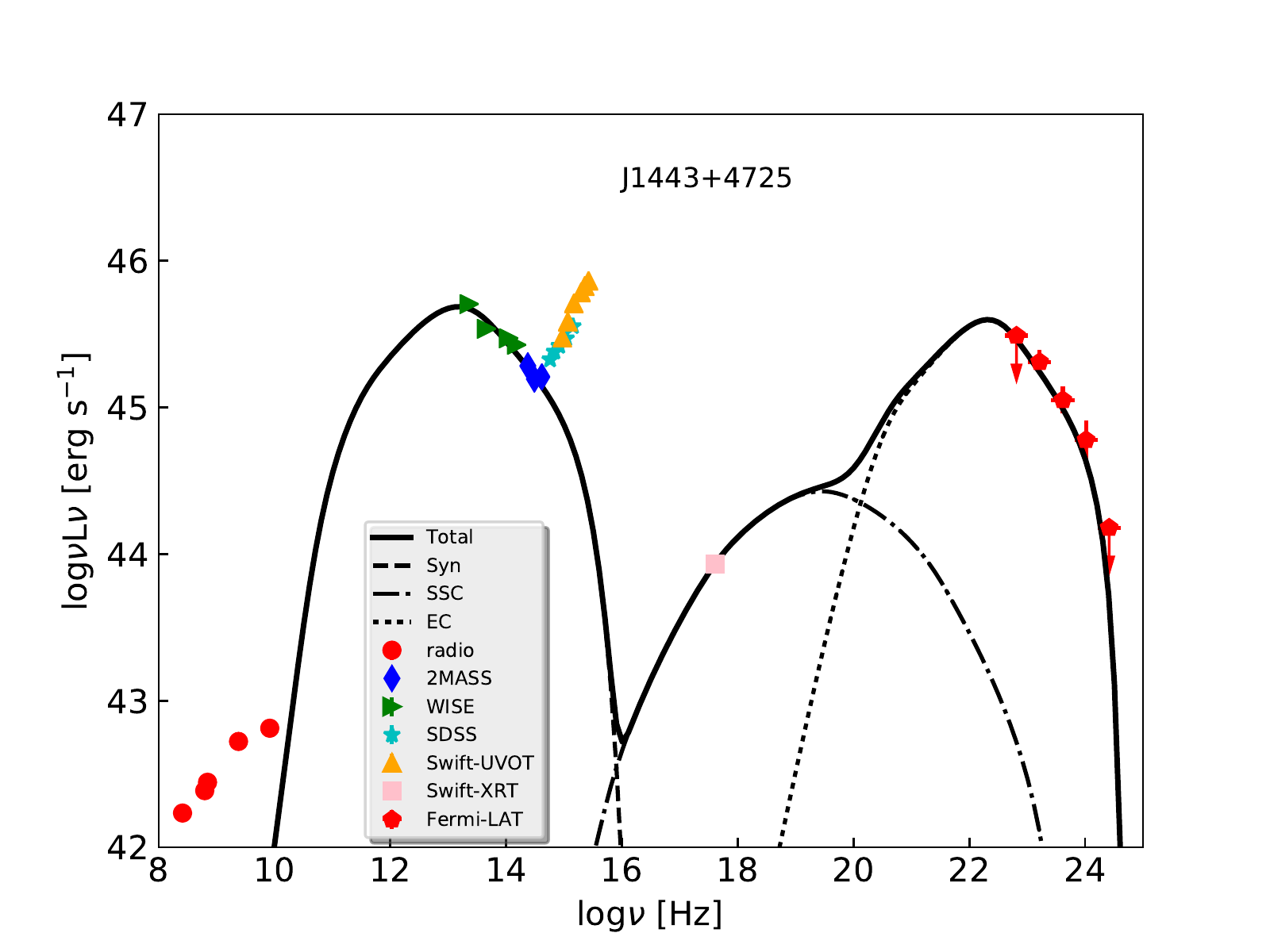}
\caption{SEDs and the jet emission modelings of J1305+5116 and J1443+4725. The radio and infrared data are derived from NED and IRSA, respectively. The SDSS magnitudes are from SDSS quasar catalogs (Schneider et al. 2005, 2007).}
\label{Fig.1}
\end{figure}

\begin{figure}
\centering
\includegraphics[scale=0.7]{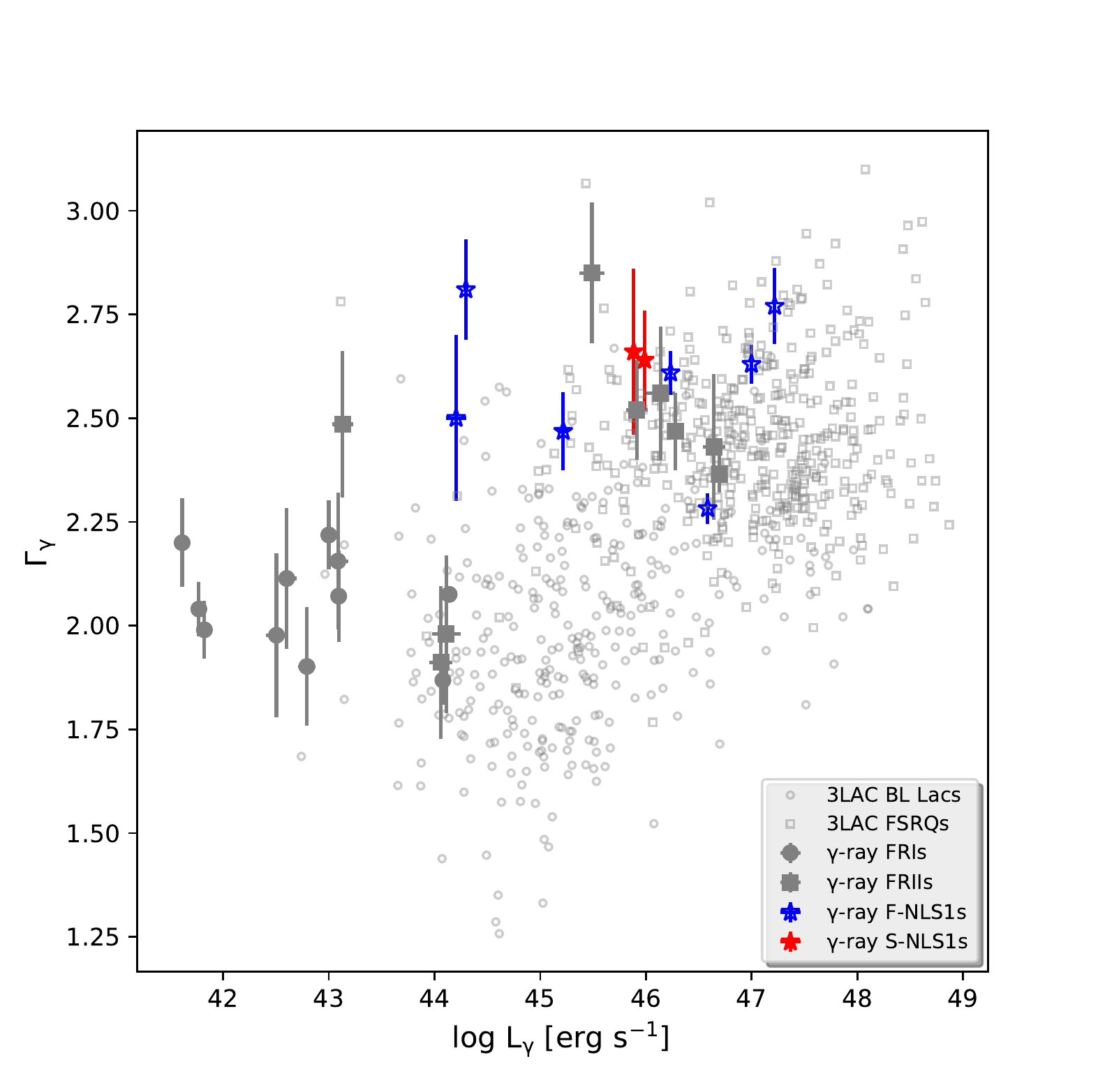}
\caption{Plane of the 0.1$-$100 GeV $\gamma$-ray luminosity and $\gamma$-ray spectral index for RLNLS1s, together with 3LAC blazars and MAGNs. $\gamma$-ray data are obtained from the ASDC data website (http://www.asdc.asi.it/fermi3lac/) except FBQS J1644+2619 whose $\gamma$-ray data is derived from D'Ammando et al.(2015). All the luminosities are evaluated in the source rest frame. }
\label{Fig.1}
\end{figure}

\begin{figure}
\centering
\includegraphics[scale=0.7]{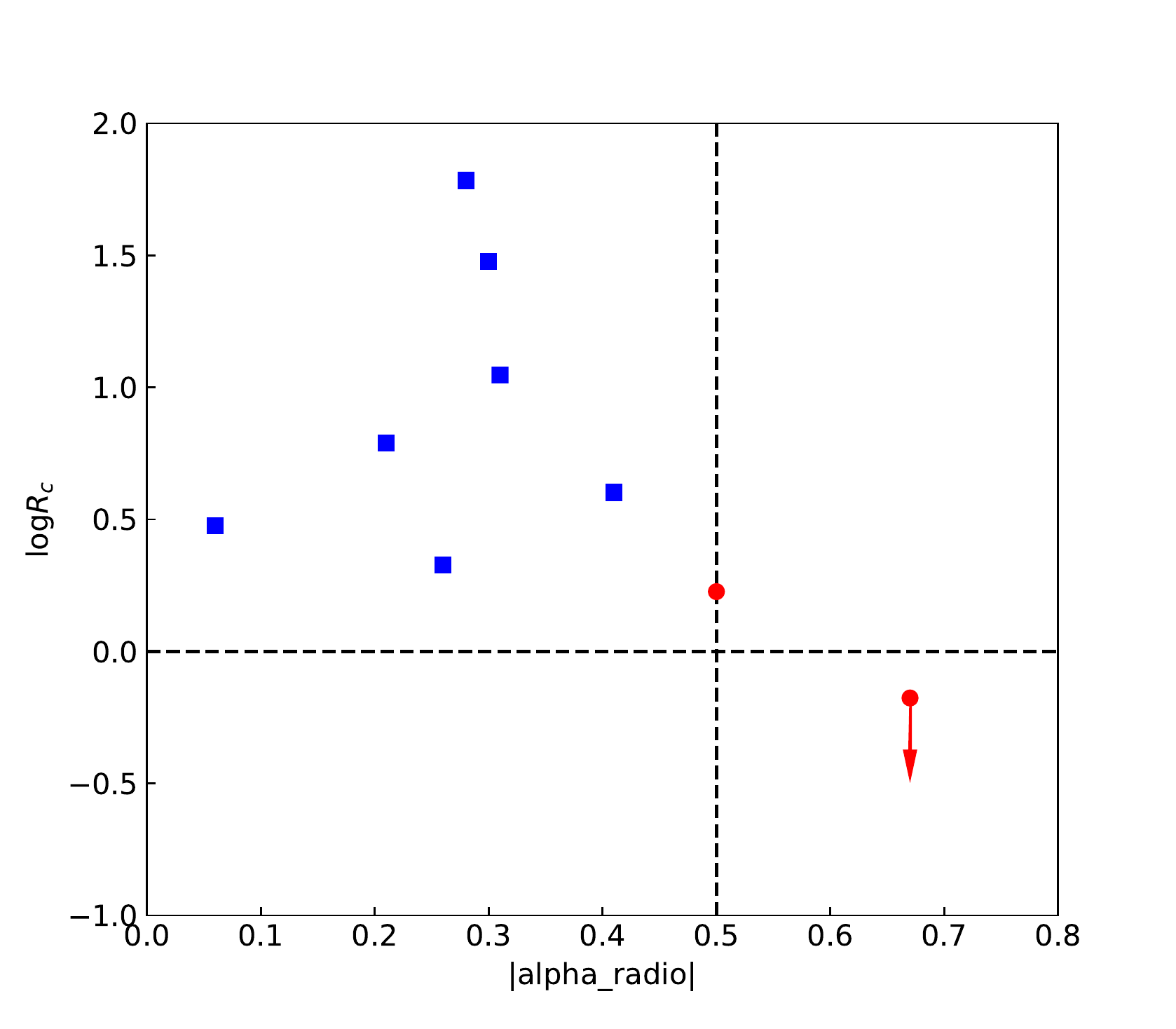}
\caption{Plane of the radio index (from 1.4 to 5 GHz) and $R_{c}$ for $\gamma$-ray NLS1s. Red circles are the two $\gamma$-ray S-NLS1s reported here, while blue squares are the known 7 $\gamma$-ray F-NLS1s. Main data of the radio index and $R_{c}$ values from 15 GHz VLBA observations are derived from Foschini et al. (2015). For FBQS J1644+2619, its $\rm R_{c}$ value is given by the 1.4 GHz VLA observation (Doi et al. 2012). For SDSS J122222.55+041315.7, its $R_{c}$ value is calculated based on the 15 GHz VLBA observation (Lister \& Homan 2005). For PKS 2004$-$447, its $R_{c}$ value is obtained from Schulz et al. (2016). The $R_{c}$ values of the S-NLS1s are derived from the 5~GHz VLBA observations (Gu et al. 2015).}
\label{Fig.1}
\end{figure}

\clearpage
\begin{deluxetable}{cccccccccccc}
\scriptsize
\tablenum{1} 
\tablewidth{0pt}
\tablecaption{Basic information and multiwavelength properties of the S-NLS1s}
\tablehead{ \colhead{Name} &\colhead{Alias NED} &\colhead{\it b} &\colhead{\it z} &\colhead{\it $\rm S_{151~MHz}$} &\colhead{$\rm S_{1.4~GHz}$} &\colhead{$m_{v}$} &\colhead{log $\rm M_{BH}$}
}

\startdata
J0146-0040 &2MASX J01464481-0040426 &$-60.4^{\circ}$ &0.083 &--- &8.7 &18.09 &7.35 \\[3pt]
J0559-5026 &PKS 0558-504 &$-28.6^{\circ}$ &0.137 &--- &--- &14.94 &7.41 \\[3pt]
J0806+7248 &RGB J0806+728 &$31.4^{\circ}$ &0.098 &137 &50.1 &--- &6.68 \\[3pt]
J0850+4626 &SDSS J085001.17+462600.5 &$39.3^{\circ}$ &0.524 &921 &16.4 &18.99 &7.65\\[3pt]
J0952-0136 &Mrk 1239 &$38.2^{\circ}$ &0.020 &--- &62.8 &14.96 &7.02 \\[3pt]
J1034+3938 &KUG 1031+398 &$59.1^{\circ}$ &0.042 &160 &25.9 &16.90 &6.3 \\[3pt]
J1200-0046 &SDSS J120014.08-004638.7 &$59.5^{\circ}$ &0.179 &--- &40.2 &17.71 &7.31 \\[3pt]
J1302+1624 &Mrk 783 &$79.0^{\circ}$ &0.067 &--- &33.2 &16.16 &7.36 \\[3pt]
J1305+5116 &SDSS J130522.74+511640.2 &$65.7^{\circ}$ &0.788 &320 &87.4 &17.22 &8.47 \\[3pt]
J1413-0312 &NGC 5506 &$53.8^{\circ}$ &0.006 &--- &339.4 &16.14 &8.07 \\[3pt]
J1432+3014 &SDSS J143244.91+301435.3 &$67.6^{\circ}$ &0.355 &333 &51.2 &18.68 &7.49\\[3pt]
J1435+3131 &SDSS J143509.49+313147.8 &$67.0^{\circ}$ &0.502 &248 &43.2 &19.34 &7.48 \\[3pt]
J1443+4725 &SDSS J144318.56+472556.7 &$60.2^{\circ}$ &0.705 &500 &166.3 &18.16 &7.36 \\[3pt]
J1450+5919 &SDSS J145041.93+591936.9 &$51.8^{\circ}$ &0.202 &--- &3.4 &18.73 &6.99 \\[3pt]
J1703+4540 &SDSS J170330.38+454047.1 &$37.5^{\circ}$ &0.060 &370 &121.8 &16.32 &6.49 \\[3pt]
J1713+3523 &FBQS J1713+3523 &$34.5^{\circ}$ &0.083 &61 &12.0 &--- &7.13 \\[3pt]
J1722+5654 &SDSS J172206.03+565451.6 &$34.5^{\circ}$ &0.426 &195 &43.0 &18.31 &7.60 \\[3pt]
J2314+2243 &RX J2314.9+2243 &$-35.0^{\circ}$ &0.169 &--- &18.7 &15.28 &7.95\\[3pt]
\enddata
\tablecomments{(1) Short name of the object; (2) alias from NED; (3) Galactic latitude; (4) redshift; (5) radio flux density at 151~MHz in mJy; (6)NVSS radio flux density at 1.4~GHz in mJy; (7) apparent $V$ band magnitude; (8) black hole mass. The Galactic latitudes and redshifts are derived from NED. The $\rm S_{151~MHz}$ is obtained from 7th Cambridge catalog (McGilchrist et al. 1990) when available, otherwise given as an extrapolation of the WNESS measurement at 352~MHz (Rengelink et al. 1997), assuming the index of radio spectrum is 0.7. For J0559-5026, it has been observed by the radio telescopes in the southern hemisphere (e.g. included in the SUMSS and ATCA catalogs). Most of the apparent $V$ band magnitudes are transformed from the SDSS $g$ and $r$ magnitudes (Schneider et al. 2007).While for J0559-5026 and J2314+2243, their magnitudes are provided in Ojha et al.(2009) and Komossa et al. (2015). The logarithm of black hole masses in scale of $M_{\odot}$ is derived from Berton et al. (2015).}
\end{deluxetable}
\clearpage

\clearpage
\begin{deluxetable}{cccccccccccc}
\scriptsize
\tablenum{2} \tablewidth{0pt}
\tablecaption{Results of 7-yr LAT data of the S-NLS1s}
\tablehead{\colhead{Name} &\colhead{TS value} &\colhead{photon flux} &\colhead{flux error} & \colhead{$\rm \Gamma_{ph}$} &\colhead{$\rm \Gamma_{ph}$ error}
}
\startdata
J1305+5116 &26.3 &4.2 &1.5 &2.7 &0.2 \\[3pt]
J1443+4725 &70.2 &6.8 &1.3 &2.6 &0.1 \\[3pt]
\hline
J0146-0040 &$<$1 &2.1 &--- &2.5f &--- \\[3pt]
J0559-5026 &2.6 &2.8 &--- &2.5f &--- \\[3pt]
J0806+7248 &$<$1 &1.0 &--- &2.5f &--- \\[3pt]
J0850+4626 &$<$1 &1.1 &--- &2.5f &--- \\[3pt]
J0952-0136 &$<$1 &1.7 &--- &2.5f &--- \\[3pt]
J1034+3938 &$<$1 &0.8 &--- &2.5f &--- \\[3pt]
J1200-0046 &5.4 &3.6 &--- &2.5f &--- \\[3pt]
J1302+1624 &1.0 &2.6 &--- &2.5f &--- \\[3pt]
J1413-0312 &6.7 &3.5 &--- &2.5f &--- \\[3pt]
J1432+3014 &$<$1 &1.3 &--- &2.5f &--- \\[3pt]
J1435+3131 &$<$1 &0.9 &--- &2.5f &--- \\[3pt]
J1450+5919 &$<$1 &0.6 &--- &2.5f &--- \\[3pt]
J1703+4540 &$<$1 &1.5 &--- &2.5f &--- \\[3pt]
J1713+3523 &$<$1 &0.8 &--- &2.5f &--- \\[3pt]
J1722+5654 &$<$1 &1.2 &--- &2.5f &--- \\[3pt]
J2314+2243 &$<$1 &1.1 &--- &2.5f &--- \\[3pt]
\enddata
\tablecomments{(1) Name of the object; (2) TS value of the 7-yr LAT data analysis; (3) $\gamma$-ray photon flux integrated between 100~MeV and 500~GeV in scale of $\rm 10^{-9}$ ph $\rm cm^{-2}$ $\rm s^{-1}$ if it is significantly detected by LAT, otherwise the 95\% C.L. upper limit is given. (4) 1-$\sigma$ statistical error of the photon flux; (5) $\gamma$-ray spectral index. For the weak sources, their index are fixed as 2.5 during the analysis; (6) error of the spectral index. Note that at this time the impact on J1305+5116 from a nearby neighbor is significant, see Section 3.1.}
\end{deluxetable}
\clearpage

\begin{deluxetable}{cccccccccccc}
\tabletypesize{\scriptsize} \tablenum{3} \tablewidth{0pt} \tablecaption{{\it Swift}-UVOT results} 

\tablehead{\colhead{Source name} & \colhead{Date} & \colhead{V} & \colhead{B} & \colhead{U} & \colhead{UVW1} & \colhead{UVM2} & \colhead{UVW2}} \startdata
J1305+5116  & 2012-06-04    & ... & ... & $17.36\pm{0.07}$ & ... & ... & ... \\
J1305+5116  & 2012-06-05 & $17.33\pm0.13$ & $17.48\pm0.10$ & $17.48\pm0.08$ & $17.57\pm0.06$ & $17.38\pm0.08$ & $17.45\pm0.07$ \\
J1305+5116  & 2012-08-21  & $17.34\pm0.13$ & $17.52\pm0.10$ & $17.48\pm0.08$ & $17.78\pm0.08$ & $17.40\pm0.08$ & $17.74\pm0.06$  \\
J1443+4725  & 2012-03-22   & ... & ... & $18.04\pm0.07$ & ... & ... & ...\\
J1443+4725  & 2012-06-02  & ... & ... & $18.07\pm0.10$ & $18.25\pm0.07$ & ... & ... \\
J1443+4725  & 2012-06-05  & ... & ... & ... & $18.22\pm0.07$ & ... & ... \\
J1443+4725  & 2012-06-16  & ... & ... & ... & ... & $18.06\pm0.09$ & ... \\
J1443+4725  & 2012-06-17  & ... & ... & $17.96\pm0.07$ & $18.21\pm0.08$ & ... & ... \\
J1443+4725  & 2012-06-19  & ... & ... & ... & ... & ... & $18.14\pm0.09$ \\
J1443+4725  & 2012-06-21  & ... & ... & ... & $18.20\pm0.09$ & ... & ... \\
J1443+4725  & 2012-06-22  & ... & ... & $18.08\pm0.06$ & ... & ... & $18.16\pm0.10$ \\
J1443+4725  & 2012-06-24  & ... & ... & ... & ... & $18.32\pm0.10$ & ... \\
J1443+4725  & 2012-06-26  &$>$17.73 & $17.97\pm0.16$ & $17.95\pm0.07$ & $18.10\pm0.12$ & $18.12\pm0.12$ & $18.19\pm0.09$ \\
J1443+4725  & 2012-06-28  & $18.00\pm0.18$ & $17.96\pm0.12$ & $17.91\pm0.09$ & $18.03\pm0.09$ & $18.07\pm0.08$ & $18.17\pm0.07$ 

\enddata

\tablecomments{The extinction corrected optical/UV magnitudes in six UVOT bands from {\it Swift} observations. The extinction corrected magnitudes are $A_{V}=0.02$, $A_{B}=0.03$, $A_{U}=0.04$,$A_{UW1}=0.05$, $A_{UM2}=0.08$, $A_{UW2}=0.07$ for J1305+5116 and $A_{V}=0.04$, $A_{B}=0.06$, $A_{U}=0.07$,$A_{UW1}=0.09$, $A_{UM2}=0.13$, $A_{UW2}=0.12$ for J1443+4725. 
\label{uvot}}
\end{deluxetable}

\begin{deluxetable}{lccccccccccc}
\scriptsize
\tablenum{4} \tablewidth{0pt}
\tablecaption{Input parameters of the SED models\tablenotemark{a}}
\tablehead{ \colhead{Source Name} &\colhead{$\rm p_{1}$} &\colhead{$\rm p_{2}$} &\colhead{$\gamma_{br}$\tablenotemark{b}} &\colhead{$\gamma_{min}$} &\colhead{K ($\rm cm^{-3}$)} &\colhead{B (Gauss)} &\colhead{$\delta$} &\colhead{$R_{b}$ (cm)}
}

\startdata
J1305+5116 &2.0 &4.0 &1024 &100 &168 &0.84 &6.6 &$\rm 2.9\times10^{17}$ \\[3pt]
J1443+4725 &2.0 &3.9 &962 &100 &204 &0.62 &6.4 &$\rm 2.9\times10^{17}$ \\[3pt]
\enddata

\tablenotetext{a}{A detailed description of the parameters is provided in the Section 5.}

\tablenotetext{b}{$\gamma_{max}$ is set as ten times of $\gamma_{br}$.}

\end{deluxetable}

\end{document}